\documentclass{svproc}
\usepackage{url}

\usepackage[dvipdfmx]{graphicx} 
\usepackage{bmpsize}

\usepackage[dvipsnames]{xcolor}
\usepackage{mathtools}

   \usepackage[makeroom]{cancel}
   \usepackage{color}

   \usepackage{array}
   \usepackage{url}

\usepackage{verbatim}
\usepackage{amsmath,bm}
\usepackage{amssymb}
\newcommand \be  {\begin{equation}}
\newcommand \bea {\begin{eqnarray} \nonumber }
\newcommand \ee  {\end{equation}}
\newcommand \eea {\end{eqnarray}}
\renewcommand{\leq}{\leqslant}
\renewcommand{\geq}{\geqslant}

\begin{document}

\mainmatter              
\title{Random Linear Systems with Quadratic Constraints: from Random Matrix Theory to replicas and back}
\titlerunning{Random Linear Systems with Quadratic Constraints}  
%
\author{Pierpaolo Vivo}
\authorrunning{Pierpaolo Vivo} 
%
\tocauthor{Pierpaolo Vivo}
\institute{Department of Mathematics\\King's College London\\
\email{pierpaolo.vivo@kcl.ac.uk}}

\maketitle              
\begin{abstract}
I present here a pedagogical introduction to the works by Rashel Tublin and Yan V. Fyodorov on random linear systems with quadratic constraints, using tools from Random Matrix Theory and replicas \cite{Yan1,Yan2}. These notes illustrate and complement
the material presented at the Summer School organised within the Puglia Summer Trimester 2023 in Bari (Italy). Consider a system of $M$ linear equations in $N$ unknowns, $\sum_{j=1}^N A_{kj}x_j=b_k$ for $k=1,\ldots,M$, subject to the constraint that the solutions live on the $N$-sphere, $x_1^2+\ldots + x_N^2=N$. Assume that both the coefficients $A_{ij}$ and the parameters $b_i$ be independent Gaussian random variables with zero mean. Using two different approaches -- based on Random Matrix Theory and on a replica calculation -- it is possible to compute whether a large linear system subject to a quadratic constraint is \emph{typically} solvable or not, as a function of the ratio $\alpha=M/N$ and the variance $\sigma^2$ of the $b_i$'s. This is done by defining a quadratic loss function $H({\bm x})=\frac{1}{2}\sum_{k=1}^M\left[\sum_{j=1}^NA_{kj} x_j-b_k\right]^2$ and computing the statistics of its minimal value on the sphere, ${\cal E}_{min}=\min_{||\bm x||^2=N}H({\bm x})$, which is zero if the system is compatible, and larger than zero if it is incompatible. One finds that there exists a {\it compatibility threshold} $0<\alpha_c<1$, such that systems with $\alpha>\alpha_c$ are typically incompatible. This means that even weakly under-complete linear systems could become typically incompatible if forced to additionally obey a quadratic constraint.
\end{abstract}

\section*{Prologue}

These lecture notes are based on a short course I gave at the Summer School organised within the Puglia Summer Trimester 2023 in Bari (Italy). There is no new research material in here. All the credit goes to Rashel Tublin and Yan V. Fyodorov, who wrote two insightful papers on the subject \cite{Yan1,Yan2}, which -- along with Rashel's Ph.D. work at King's College London \cite{tublin} -- I have here drawn liberally and \emph{verbatim} from. I wish to thank both of them for writing such clear papers and for allowing me access to their deep knowledge and insight about this problem. I take this opportunity to warmly thank all the organisers of the Summer School -- Biagio Cassano, Fabio Deelan Cunden, Matteo Gallone, Marilena Ligab\`o, Alessandro Michelangeli, and Domenico Pomarico -- for the flawless organisation, for putting together such a rich and diverse programme, and for many inspiring conversations. I cherish their friendship dearly. I would also like to thank my fellow speakers at the School, Nilanjana Datta and Riccardo Adami, from whom I learnt so much in such a short period of time. I enjoyed their company immensely. Finally, I extend my gratitude to all the students and early-career researchers who participated in the School for their engagement and the interest they showed in what I had to say.

I chose to prepare my short course on the topic of optimisation on the sphere as it allowed me the rare luxury of being able to present the same calculation -- the average minimal loss, signalling typical compatibility/incompatibility of a random linear system of equations with a quadratic constraint -- in two alternative ways: one based on the synergy between Lagrange optimisation and Random Matrix Theory, the other based on the heuristic ``replica method'' from the physics of disordered systems. I tried to explain all the steps in full detail and make these notes as self-contained -- and hopefully instructive -- as possible. 

\section{Introduction}

{\bf Scenario 1.}
Consider the following linear system of two equations ($M=2$) in three unknowns ($N=3$)
\begin{align}
x_1+x_2-x_3 &=b_1\\
x_1+2x_2+x_3 &=b_2\ ,
\end{align}
for $b_1,b_2$ real parameters.
Clearly, any triple of the form\\ $\left(2b_1-b_2+3z,-b_1+b_2-2z,z\right)$ with $z\in\mathbb{R}$ is a solution of the system.

What happens if we add the constraint that the solution vector $\bm x = (x_1,x_2,x_3)$ must live on the sphere of square radius $r^2=N=3$, namely 
\begin{equation}
x_1^2+x_2^2+x_3^2=3\ ?
\end{equation}
The system now only admits real solutions if
\begin{equation}
-3\leq b_1\leq 3\qquad\text{and}\qquad \frac{2 b_1}{3}-\frac{1}{3} \sqrt{14} \sqrt{9-b_1^2}\leq b_2\leq \frac{1}{3} \sqrt{14} \sqrt{9-b_1^2}+\frac{2 b_1}{3}\ ,
\end{equation}
i.e. for ``small enough'' known coefficients. This simple example shows that adding a quadratic constraint to a linear system that would otherwise have infinitely many solutions may turn it into an incompatible system if the known values of $b_i$ in the right-hand sides of linear equations are ``large enough''.\\
\\
{\bf Scenario 2.}
Consider now the following \emph{oblique Procrustes problem}:\\
\\
{\it Given a $M \times N, \, M>N$ matrix $A$ and a ``target structure'' matrix $B$ of the same dimension, find a $N\times N$ matrix $X$ such that 
\begin{equation}
B=AX \label{Procr}
\end{equation}
holds with maximal precision, and columns of $X$ are of unit norm}.\\
\\
We ask the following questions:
\begin{enumerate}
\item Is there a relation between the two scenarios?
\item What is the significance of these problems?
\end{enumerate}

\noindent ``Matrix fitting'' problems as the one described in Scenario 2 fall under the umbrella label of {\bf Procrustes problems}. According to Greek mythology, Procrustes, son of Poseidon, was a rogue smith and a bandit, who had a bed in his stronghold, in which he invited every passer-by to spend the night. Using his smith's hammer, he would then get to work on his guests to stretch or mutilate them until they fit his bed. Nobody ever fit the bed exactly. The ``target structure'' matrix $B$ in Eq. \eqref{Procr} plays the role of Procrustes' bed, the matrix $A$ of the passer-by, and $X$ of the smith's hammer.

A prominent application of the oblique Procrustes problem is in the context of multiple \emph{factor analysis}, a field of psychological research pioneered by Spearman \cite{spearman} and Thurstone \cite{thurstone1,thurstone2}. In quite general terms, the correlation between a number of observed variables in an experiment or test -- say, the exam scores by $N$ students in $M$ subjects -- may be explained by a much smaller (but unobserved) number of variables (the \emph{factors}), for example different types of ``intelligence'' (scientific, linguistic, social etc.). Once the number of factors are posited or guessed based on external evidence, one seeks for the \emph{loading matrix} that best describes the data, i.e. the makeup and weighted combination of intelligence traits that best defines each student given their exam results. It turns out that -- under quite general assumptions -- the problem of determining the best loading matrix can be cast in the oblique Procrustes form given above.

\begin{figure}[htb]
\begin{center}
\includegraphics[scale=0.25,clip=true]{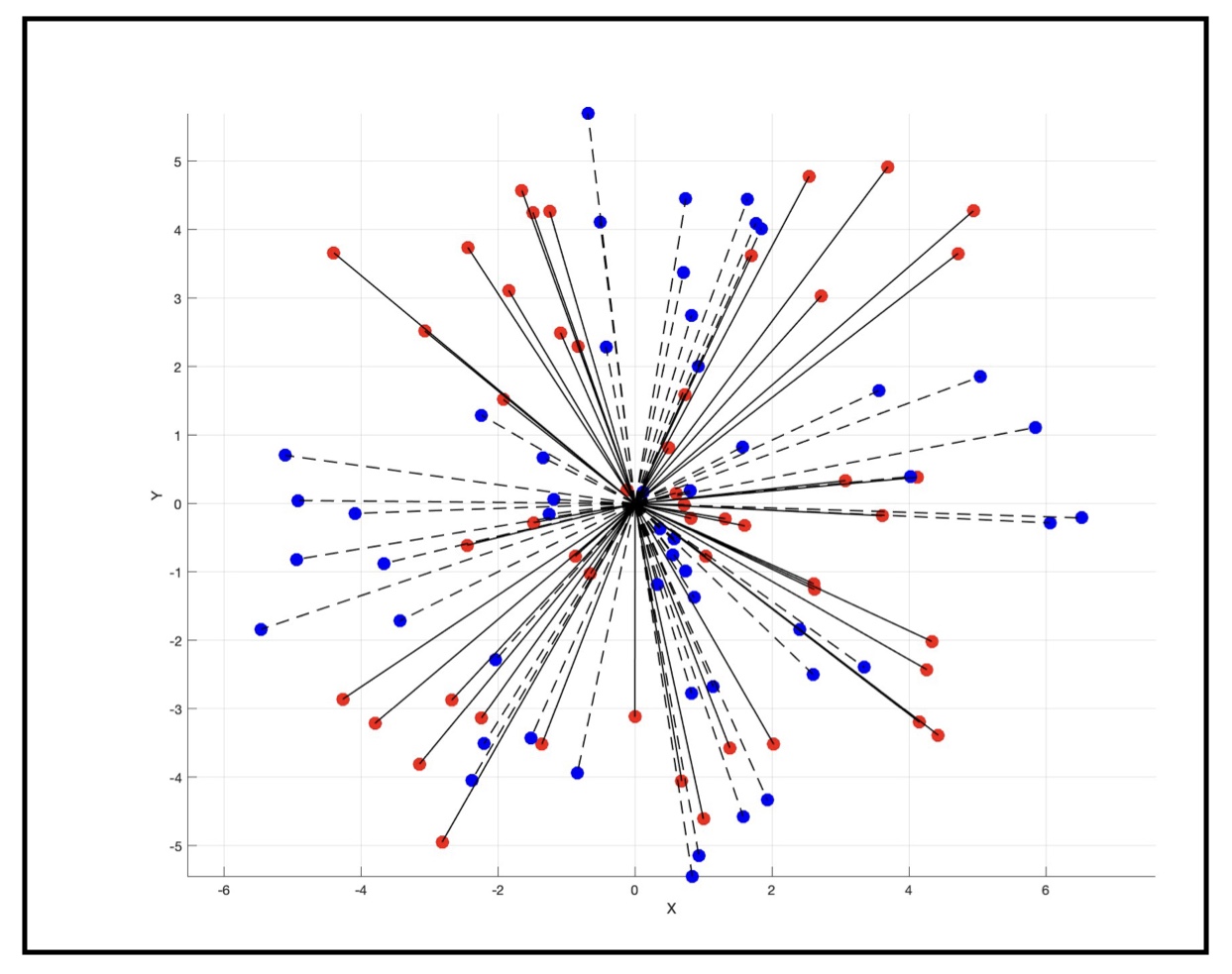}
\caption{Setting of the orthogonal Procrustes problem. A cloud of red points is first rigidly rotated clockwise via an orthogonal matrix $\Omega$, and some small Gaussian noise is then added to the resulting rotated points, yielding the final blue points. Given the two clouds (red and blue points), one seeks for the ``best'' orthogonal matrix $\tilde\Omega$ that would bring the blue points ``as close as possible'' to the initial red points.} \label{figprocr}
\end{center}
\end{figure}

Another Procrustes problem -- the so-called \emph{orthogonal} version --- with applications in computer vision \cite{compvision}, analytical chemistry \cite{chem}, speech translation \cite{linguistics1,linguistics2} and many others, seeks for the ``best'' orthogonal matrix $\tilde\Omega$ that connects two clouds of points (red and blue) on the plane, where the blue points have been first rigidly rotated with respect to the red ones, and then smeared on the plane by adding some small Gaussian noise (see Fig. \ref{figprocr} for an illustration, and check \cite{orthprocrustes} for a very pedagogical explanation of the setting, the analytical solution, and a Julia computer code to implement the analytical solution).  

Coming back to the \emph{oblique} Procrustes problem described in Scenario 2, we may write the system of equations \eqref{Procr} for the columns $\{\bm x\}$ of X and $\{\bm b\}$ of $B$ -- and set the norm of $\bm x$ to $N$ instead of $1$ for later convenience. This way,  we can reformulate the oblique Procrustes problem by asking for a solution $\bm x$ -- a $N$-dimensional vector -- of the over-complete linear system
\begin{equation}
A\bm x = \bm b\ ,\label{originalsystem}
\end{equation}
with $A$ a $M\times N$ matrix, supplemented with the condition $||\bm x||^2=\bm x^T\bm x=N$, where $||\bm x||^2=x_1^2+\ldots + x_N^2$ is the squared norm of the vector, and $(\cdot)^T$ denotes vector/matrix transposition. This way, we cast the problem described in Scenario 2 precisely in the form of Scenario 1.

As is easy to see, the system of equations in \eqref{originalsystem} is over-complete for $M>N$ -- i.e., it has more equations than unknowns -- so in general, one cannot find a vector $\bm x$ that satisfies it exactly. To find an approximate solution, the best one can do is to minimise some cost function that penalises deviations from the relation $A\bm x = \bm b$, where $\bm x$ and $\bm b$ denote the corresponding columns of $X$ and $B$ respectively. In this setting, the \emph{least square fitting} is one of the most natural and frequently used minimisations, corresponding to the \emph{cost function}
\begin{equation}
H(\bm x)=\frac{1}{2}||A\bm x - \bm b||^2\ ,\label{defHx}
\end{equation}
to be minimised over the sphere $||\bm x||^2=N$. If a vector $\bm x$ with squared norm equal to $N$ can be found for which $H(\bm x)=0$, then the system \eqref{originalsystem} is precisely compatible, with solution $\bm x$. If this is not possible, then $\bm x = \textrm{argmin}~H(\bm x)$ -- i.e. the vector that minimises $H(\bm x)$ on the sphere -- is the best approximation (in the sense of squared norm) that we could find for the incompatible linear system \eqref{originalsystem}. This problem attracted considerable attention starting from the work \cite{Browne1967}, see e.g. \cite{Gander1981,GolubMatt1991}.

So far, we have not specified the matrix $A$ and the vector of coefficients $\bm b$ forming the linear system \eqref{originalsystem} that we look for an exact solution of - or the best approximation to. And we are not going to do that!

The papers \cite{Yan1,Yan2} indeed analyse a \emph{random} version of the Procrustes problem, where both the matrix $A$ and the vector $\bm b$ are drawn at random from a Gaussian probability density function. The reason for doing this is to analyse \emph{average/typical} properties of the linear system \eqref{originalsystem} supplemented with a quadratic constraints, which do not depend on the fine details of individual instances of the problem.

More precisely, consider the $M\times N$ entries $A_{ij}$ to be independently drawn at random from a Gaussian probability density function with mean zero, and variance $1/N$,
\begin{equation}
\langle A_{ij}\rangle_A=0\qquad \langle A_{kj}A_{\ell m}\rangle_A=(1/N)\delta_{k\ell}\delta_{jm}\ ,\label{varianceA}
\end{equation}
while the $M$ entries $b_k$ of the vector $\bm b$ are independently drawn from a Gaussian probability density function with mean zero, and variance $\sigma^2$,
\begin{equation}
\langle b_k\rangle_{\bm b}=0\qquad \langle b_k b_\ell\rangle_{\bm b}=\sigma^2\delta_{k\ell}\ ,
\end{equation}
where
\begin{equation}
\langle \cdot\rangle_A = \int \mathrm{d}A~P_A(A)(\cdot)
\end{equation}
denotes averaging over the joint probability density (jpd)\\ $P_A(A):=P_A(A_{11},\ldots,A_{MN})$ of the matrix $A$ (and similarly for the vector $\bm b$).

The cost function
	\begin{equation}\label{energy}
		H({\bm x})=\frac{1}{2}||A{\bm x}-{\bm b}||^2:=\frac{1}{2}\sum_{k=1}^M\left[\sum_{j=1}^NA_{kj} x_j-b_k\right]^2
	\end{equation}
is therefore a random variable, as it depends on the realisation of the randomness (\emph{disorder}) in $A$ and $\bm b$. We are interested in the statistics of the minimal cost
\begin{equation}
 {\cal E}_{min}=H\left({\bm x}_{min}\right)=\min_{||\bm x||^2=N} H(\bm x)\ ,
\end{equation}
corresponding to the ``best'' approximation to the solution of the linear system \eqref{originalsystem} on the sphere, for large $N,M$, as a function of the parameters $\alpha=M/N$ (ratio between number of equations and number of unknowns) and $\sigma^2$ (variance of the ``noise'' term).

Following \cite{Yan1} and \cite{Yan2}, to achieve this goal we will use two different (and largely complementary) approaches. The first is based on the method of Lagrange multipliers combined with tools from Random Matrix Theory (RMT) and is developed in Section \ref{sec:Lagrange} for the case $M,N\to\infty$ with $\alpha=M/N>1$. The outcome of this calculation is a precise formula (see Eq. \eqref{finalEminLagrange}) for
\begin{equation}
\lim_{N,M\to\infty} \frac{\langle  {\cal E}_{min}\rangle_{A,\bm b}}{N}\ ,
\end{equation}
where the average $\langle \cdot\rangle_{A,\bm b}$ is taken with respect to the Gaussian disorder in $A$ and $\bm b$. While derived only in the regime $\alpha>1$, the formula \eqref{finalEminLagrange} confirms the natural expectation that over-complete systems with an additional quadratic constraint are typically incompatible, but hints towards the existence of an intermediate regime for ``weakly'' under-complete systems ($0<\alpha_c<\alpha<1$) that are also typically incompatible. This is at odds with what happens without spherical constraint, where under-complete systems $\alpha<1$ are always typically compatible (see Appendix \ref{appnospherical}).

A second method -- used in Section \ref{sec:replicas} -- exploits ideas from Statistical Mechanics for directly searching and characterising ${\cal E}_{min}$. When doing this one can dispose of the restriction $\alpha=M/N>1$. The method is based on setting up an auxiliary thermodynamical system of particles subject to a quadratic potential and in equilibrium at inverse temperature $\textcolor{red}{\beta}$: the zero-temperature free energy of the system can be easily related to ${\cal E}_{min}$. Taking the average over the disorder in $A$ and $\bm b$ is the main technical challenge, as it involves a so-called \emph{quenched} average of the logarithm of the associated partition function of the system. The technical challenge can be overcome using a heuristic method -- developed in the field of disordered systems and spin glasses -- called the \emph{replica method}.

\section{Lagrange multiplier method}\label{sec:Lagrange}

We start with presenting how to set up the Lagrange multiplier method assuming $M>N$ (over-complete system, with more equations than unknowns).
 Following the standard idea of a constrained minimisation, one uses the cost function \eqref{energy} to build the associated Lagrangian 
 \begin{equation}
 {\cal L}_{\lambda}({\bm x})=H({\bm x})-\frac{\lambda}{2}{\bm x}^T{\bm x}=\frac{1}{2}\sum_{k=1}^M \left[\sum_{j=1}^N A_{kj}x_j-b_k\right]^2-\frac{\lambda}{2}\sum_k x_k^2\ ,
 \end{equation}
 with $\lambda\in\mathbb{R}$ being the Lagrange multiplier taking care of the spherical constraint. 
 
  The stationarity conditions $\nabla {\cal L}_{\lambda}({\bm x})=0$ entails computing the partial derivative $\partial_{x_r} {\cal L}_{\lambda}({\bm x})$ w.r.t. the component $x_r$, which reads explicitly
  \begin{equation}
  \frac{1}{2}\times 2\sum_k  \left[\sum_{j=1}^N A_{kj}x_j-b_k\right]A_{kr}-\frac{\lambda}{2}\times 2 x_r = 0\ .
  \end{equation}
 This equation can be cast in the form
 \begin{equation}
 A^T [A\bm x-\bm b]=\lambda \bm x\Rightarrow A^T A \bm x -A^T\bm b=\lambda\bm x\Rightarrow (A^T A-\lambda\mathbb{I}_N)\bm x=A^T\bm b\ ,\label{Lagr1}
 \end{equation}
 where $\mathbb{I}_N$ is the $N\times N$ identity matrix.
 
 Solving for $\bm x$, we get
 \begin{equation}
\boxed{ \bm x = (W-\lambda\mathbb{I}_N)^{-1} A^T\bm b}\ ,\label{defX}
 \end{equation}
 where $W=A^T A$ is the so-called \emph{Wishart} matrix of size $N$ -- see Appendix \ref{app:Wishart} for generalities on the Wishart ensemble.
 
 Imposing now the spherical constraint $\bm x^T \bm x = N$ yields from Eq. \eqref{defX}
 \begin{equation}
\boxed{	{\bm b}^TA(W-\lambda \mathbb{I}_N)^{-2}A^T{\bm b}=N}\ ,\label{sphconstraint}
 \end{equation}
 where we used the transposition rule for matrices $(AB)^T=B^TA^T$.
 
 The way to interpret this equation is as follows: fix an instance $A$ from the random matrix ensemble, and an instance $\bm b$ from the random vector ensemble, then find all the real solutions $\{\lambda\}$ of Eq. \eqref{sphconstraint}. Every such solution is an acceptable Lagrange multiplier corresponding to a stationary point\footnote{This stationary point may be a minimum, a maximum, or a saddle-point, though.} of the cost function.  Then, insert each of these Lagrange multipliers into Eq. \eqref{defX} and compute the corresponding vector $\bm x_\lambda$, which provides the ``location'' of the corresponding stationary point of $H(\bm x)$. An important property proved in \cite{Browne1967} (see Appendix \ref{appBrowne} for a sketch of the proof) is that the order of Lagrange multipliers exactly corresponds to the order of values taken by the cost function at the corresponding  point ${\bm x}$.  Namely, denoting $\cal N$ the total number of Lagrange multipliers, assumed to be distinct and ordered as  $\lambda_1<\lambda_2<\ldots <\lambda_{\cal N}$, such order implies $H({\bm x}_1)<H({\bm x}_2)<\ldots <H\left({\bm x}_{\cal N}\right)$. Thus the minimal loss\footnote{Since the $N$-sphere is compact and $H(\bm x)$ is continuous, it follows from the Extreme Value Theorem that the absolute minimum exists and is reached. Given Browne's theorem, $\lambda_1$ must therefore correspond to the absolute minimum (not to a saddle, or to a maximum).} is always given by 
 \begin{equation}
 {\cal E}_{min}=H\left({\bm x}_{min}\right)\ ,\label{defEmin}
 \end{equation}
 where ${\bm x}_{min}$ corresponds to $\lambda_1:=\lambda_{min}$.

 To develop some intuition about the number of solutions of Eq. \eqref{sphconstraint} as a function of the noise variance parameter $\sigma$, it is useful to write another representation of the matrix 
 \begin{equation}
 A(W-\lambda \mathbb{I}_N)^{-2}A^T\label{startEq}
 \end{equation}
 appearing on the left-hand side of Eq. \eqref{sphconstraint} which turns out to be more insightful. To this end, alongside with the Wishart $N\times N$ matrix  $W = A^T A$, we define the associated \emph{anti-Wishart} $M \times M$ matrix $W^{(a)}=AA^T$ \cite{antiwishart1,antiwishart2}. Recalling $M> N$, the two matrices share the same set of nonzero eigenvalues $\{s_i\}_{i=1}^N$ and $W^{(a)}$ has another set of $M-N$ eigenvalues equal exactly to $0$ (see Appendix \ref{app:Wishart} for a proof). Note that  $\{s_i\}_{i=1}^N$ are all non-negative, as $W$ is a correlation matrix (symmetric and positive semi-definite), and the typical scale of the Wishart eigenvalues -- if the variance of the entries of $A$ is $\sim\mathcal{O}(1/N)$ as per Eq. \eqref{varianceA} -- is $\sim\mathcal{O}(1)$.
 
 The main ingredients are:
 \begin{enumerate}
 \item The formal ``geometric series'' expansion 
 \begin{equation}
 (W-\lambda \mathbb{I}_N)^{-2}=\sum_{k=0}^\infty \frac{(k+1)W^k}{\lambda^{k+2}}\ ,
 \end{equation}
 which can be proven by differentiating w.r.t. $\mu$ the formal identity 
 \begin{equation}
 (\mathbb{I}_N-\mu W)^{-1}=\sum_{k=0}^\infty \mu^k W^k\ ,
 \end{equation}
 and setting $\mu = 1/\lambda$ at the end.
  \item the rewriting
 \begin{equation}
 A W^k A^T = \underbracket{A(A^T} \underbracket{A)(A^T} A)\cdots (A^T \underbracket{A)A^T}=[W^{(a)}]^{k+1}\ ,\label{grouping}
 \end{equation}
 obtained by grouping the matrices differently.
 \item the spectral decomposition $W^{(a)}=\sum_{i=1}^N s_i \bm v_i\bm v_i^T$ where $\{\bm v_i\}_{i=1}^M$ are the normalised eigenvectors of $W^{(a)}$ corresponding to its non-zero eigenvalues $s_i, \, i=1, \ldots, N$, as well as its corollary 
 $\varphi (W^{(a)})=\sum_{i=1}^N \varphi(s_i) \bm v_i\bm v_i^T$ for any analytic function $\varphi(z)$.  
 \end{enumerate}

Starting from Eq. \eqref{startEq}, we have
 \begin{align}
 & \nonumber A(W-\lambda \mathbb{I}_N)^{-2}A^T\underbrace{=}_{\stackrel{\uparrow}{(1.)}}\sum_{k=0}^\infty \frac{(k+1)AW^k A^T}{\lambda^{k+2}}\underbrace{=}_{\stackrel{\uparrow}{(2.)}}\sum_{k=0}^\infty \frac{(k+1)[W^{(a)}]^{k+1}}{\lambda^{k+2}}\\
  &=W^{(a)}(\lambda\mathbb{I}_M-W^{(a)})^{-2}\underbrace{=}_{\stackrel{\uparrow}{(3.)}}\sum_{i=1}^N\frac{s_i}{(\lambda-s_i)^2}\bm v_i\bm v_i^T\ .\label{der2}
 \end{align}
 
 Sandwiching both sides between ${\bm b}^T$ and ${\bm b}$ (see Eq. \eqref{sphconstraint}), we obtain the spherical constraint equality in the form
 \begin{equation}
 \sum_{i=1}^N\frac{s_i}{(\lambda-s_i)^2}({\bm b}^T\bm v_i)(\bm v_i^T {\bm b})=N\Rightarrow \boxed{ \sum_{i=1}^N\frac{s_i}{(\lambda-s_i)^2}(\bm\xi^T\bm v_i)^2=\frac{N}{\sigma^2}}\ ,\label{spherical2}
 \end{equation}
 where we set $\bm b =\sigma^2\bm\xi$, with $\bm\xi$ having standard Gaussian entries (zero mean and variance one). 
 
\begin{figure}[htb]
\begin{center}
\includegraphics[scale=0.3,clip=true]{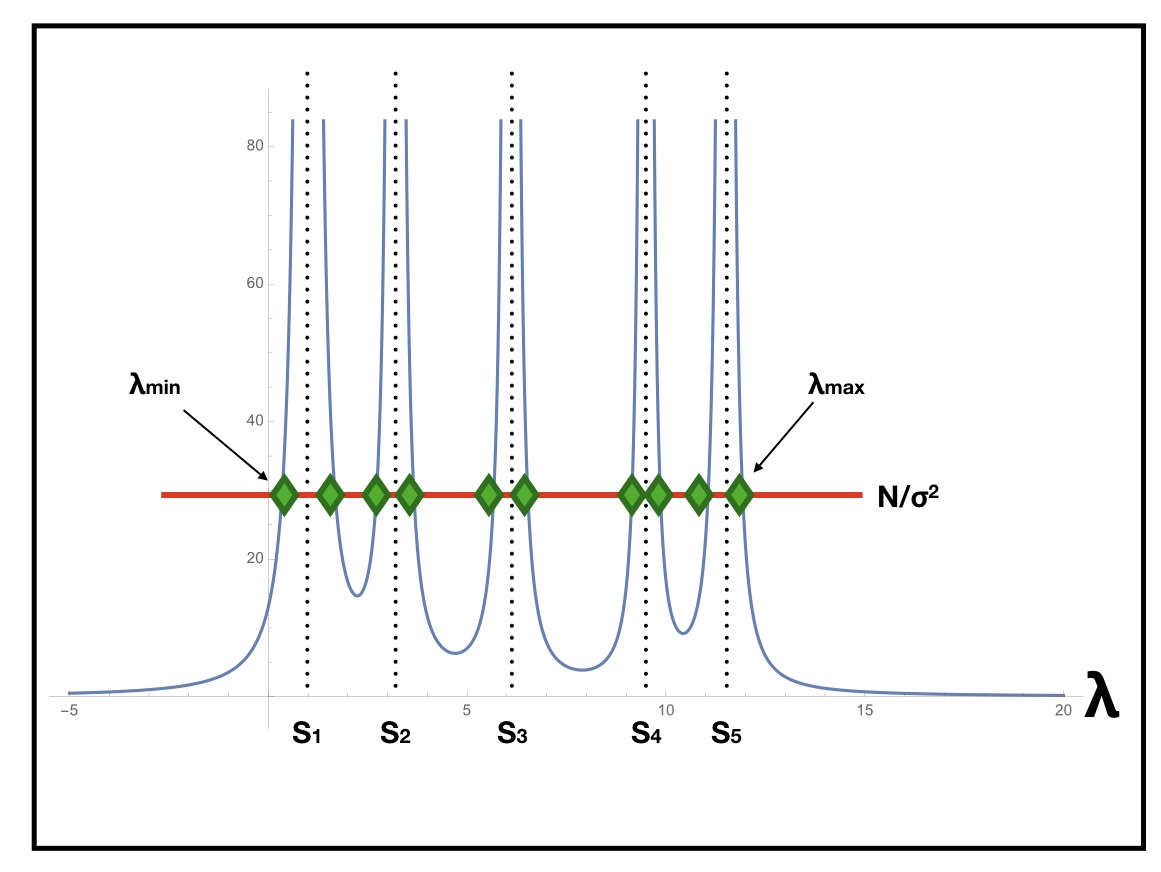}
\caption{Sketch of the plot of the left hand side of Eq. \eqref{spherical2} as a function of $\lambda$ for $N=5$. The horizontal red line signals the right hand side constant value $N/\sigma^2$, and green points highlight the intersections (solutions of the equation \eqref{spherical2}). The leftmost intersection (to the left of the smallest eigenvalue $s_1$) corresponds to the minimal cost.} \label{figG}
\end{center}
\end{figure}

It is easy to see that the left-hand side is a positive function of $\lambda$ having a single minimum between every consecutive pair of eigenvalues of $W$, see figure \ref{figG} for $N=5$ below. 

This implies there are typically 0 or 2 solutions of eq.~\eqref{spherical2} (and 1 solution with probability zero at exceptional points) for $\lambda$ between every consecutive pair of eigenvalues, plus two more solutions: the minimal one $\lambda_{min}\in (-\infty, s_1)$ and the maximal one $\lambda_{max}\in ( s_N, \infty)$. Note that the latter two solutions exist for any value of $\sigma\in [0,\infty]$, whereas by changing $\sigma$ one changes the number of solutions available between consecutive eigenvalues. In particular, in the limit of vanishing noise (i.e. $\sigma\to 0$, hence ${\bm b}\to 0$), every stationary point solution for the Lagrange multiplier corresponds to an eigenvalue $s_k$ of the Wishart matrix, with ${\bm x}=\pm {\bm e}_k$ being the associated eigenvectors (hence there are $2N$ stationary points).
	On the other hand when $\sigma \to \infty$ the ratio $N/\sigma^2$ in the right-hand side becomes smaller than the global minimum of the left-hand side in $[s_1,s_N]$. Then  only two stationary points survive: $\lambda_N:=\lambda_{max}$ and $\lambda_1:=\lambda_{min}$ -- which correspond to the global maximum and the global minimum of $H(\bm x)$.
	
Combining Eq. \eqref{defEmin} with Eq. \eqref{energy} and Eq. \eqref{defX}, we have
\begin{equation}
 {\cal E}_{min}=H\left({\bm x}_{min}\right) = H\left((W-\lambda_1\mathbb{I}_N)^{-1} A^T\bm b\right)=\frac{1}{2}\Big|\Big|A(W-\lambda_1\mathbb{I}_N)^{-1} A^T\bm b-\bm b\Big|\Big|^2\ ,\label{defEmin2}
\end{equation}
where $\lambda_1$ is the minimal solution of Eq. \eqref{spherical2}, and $W=A^T A$. We stress that $ {\cal E}_{min}$ is a random variable, which depends on the realisation of the disorder in the matrix $A$ and the parameters $\bm b$ both directly, and through the minimal value of the Lagrange multiplier $\lambda_1$.

To compute the average minimal loss, $\langle  {\cal E}_{min}\rangle_{A,\bm b}$, we proceed in two steps: (i) we compute the average/typical value $\lambda^\star$ of $\lambda_1$, and we replace the random variable $\lambda_1$ with $\lambda^\star$ in \eqref{defEmin2}. Then (ii) we compute the average $\langle  {\cal E}_{min}\rangle_{A,\bm b}$, where any indirect dependence on $A,\bm b$ has been removed. This two-step procedure of course hinges on a \emph{self-averaging} assumption, whereby the random variable $\lambda_1$ quickly converges to (and is therefore well approximated by) its mean value $\lambda^\star$.

\subsection{Calculation of $\lambda^\star$}

We show in this section that the average value $\lambda^\star$ of the smallest solution $\lambda_1$ of Eq. \eqref{spherical2} is given by
\begin{equation}
\boxed{\lambda^\star=\frac{(\sqrt{\alpha}-\sqrt{1+\sigma^2})(\sqrt{\alpha(1+\sigma^2)}-1)}{\sqrt{1+\sigma^2}}}\ ,\label{lambdastar}
\end{equation}
in the limit $N,M\to\infty$ with $\alpha=M/N$ fixed -- where $M$ is the number of equations and $N$ the number of unknowns -- and $\sigma^2$ is the variance of the noise terms $b_k$. 

We arrive at Eq. \eqref{lambdastar} starting from Eq. \eqref{spherical2}, which is rewritten as
\begin{equation}
g(\lambda)=\frac{1}{N}\sum_{i=1}^N\frac{s_i}{(\lambda-s_i)^2}(\bm\xi^T\bm v_i)^2=\frac{1}{\sigma^2}\ .\label{equatingglambda}
\end{equation}
We wish to take the average $\langle g(\lambda)\rangle_{A,\bm \xi}$ and find the minimal solution of $\langle g(\lambda)\rangle_{A,\bm \xi}=1/\sigma^2$ for large $N,M$.

First, taking the average over the variables $\bm\xi$ reads
\begin{equation}
\langle g(\lambda)\rangle_{\bm \xi}=\frac{1}{N}\sum_{i=1}^N\frac{s_i}{(\lambda-s_i)^2}\sum_{\ell,m}\underbrace{\langle \xi_\ell\xi_m\rangle_{\bm\xi}}_{\delta_{\ell m}} v_{i\ell}v_{im}=
\frac{1}{N}\sum_{i=1}^N\frac{s_i}{(\lambda-s_i)^2}\underbrace{\sum_{\ell}v_{i\ell}^2}_{=1}\ ,\label{glambda29}
\end{equation}
where we first used that $\bm\xi$ are i.i.d. standard Gaussian variables, and then the normalisation of eigenvectors of $W^{(a)}$.

\begin{figure}[htb]
\begin{center}
\includegraphics[scale=0.25,clip=true]{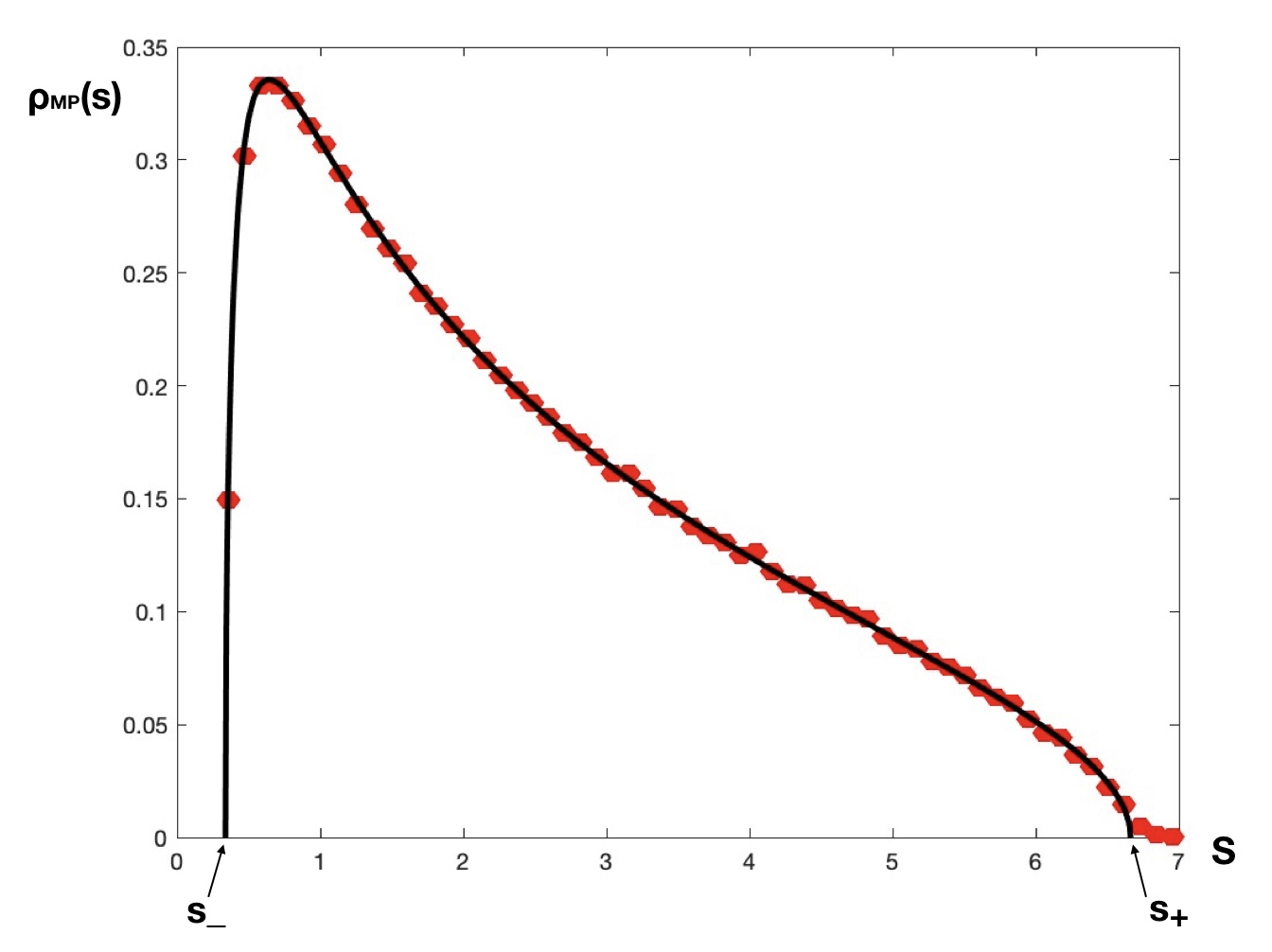}
\caption{ Red points: histogram of eigenvalues $s_i$ of a Wishart ensemble generated from $500$ ``data'' matrices $A$ each of size $M\times N$, with $M=500$, $N=200$ (corresponding to $\alpha=2.5$) and filled with i.i.d. Gaussian entries with mean zero and variance $1/N$. Solid black line: the Mar\v cenko-Pastur law \eqref{MPeq} over the compact support $[s_-,s_+]$, with $s_-=(\sqrt{\alpha}-1)^2\approx 0.338$ and $s_+=(\sqrt{\alpha}+1)^2\approx 6.662$, showing excellent agreement.} \label{figMP}
\end{center}
\end{figure}

Next, we use the property
\begin{equation}
\lim_{\stackrel{N,M\to\infty}{M/N=\alpha}}\Big\langle\frac{1}{N}\mathrm{Tr}~\varphi(W)\Big\rangle_A =\lim_{\stackrel{N,M\to\infty}{M/N=\alpha}}\Big\langle\frac{1}{N}\sum_{i=1}^N\varphi(s_i)\Big\rangle_A = \int_{s_-}^{s_+}\mathrm{d}s~\rho_{MP}(s)\varphi(s)\ ,\label{traceproperty}
\end{equation}
where $\{s_i\}$ are the eigenvalues of the $N\times N$ Wishart matrix $W=A^T A$, and 
\begin{equation}
\rho_{MP}(s)=\frac{2}{\pi s}\frac{1}{(\sqrt{s_+}-\sqrt{s_-})^2}\sqrt{(s_+-s)(s-s_-)}\label{MPeq}
\end{equation}
is the \emph{Mar\v cenko-Pastur} probability density function\footnote{It is easy to check that $\int_{s_-}^{s_+}\mathrm{d}s~\rho_{MP}(s)=1$.} \cite{MP}, having compact support between the edge points $s_\pm=(\sqrt{\alpha}\pm 1)^2$ on the positive real axis. In Fig. \ref{figMP}, we test numerically the Mar\v cenko-Pastur law on randomly generated Wishart matrices.

The lower edge point $s_-$ corresponds to the average location of the smallest eigenvalue of the Wishart matrix in the large $N,M$ limit. Since we are interested in the smallest intersection $\lambda^\star$ of the function $g(\lambda)$ with the horizontal line $1/\sigma^2$, we should be looking for this solution to the left of $s_-$ (see Fig. \ref{figG}). 

Combining \eqref{glambda29} and \eqref{traceproperty}, we have that
\begin{equation}
\lim_{\stackrel{N,M\to\infty}{M/N=\alpha}}\langle g(\lambda)\rangle_{A,\bm \xi}= \int_{s_-}^{s_+}\mathrm{d}s~\rho_{MP}(s)\frac{s}{(\lambda-s)^2}\ .\label{intMPphi}
\end{equation}
This integral is computed in Appendix \ref{app:integralMP} to give
\begin{equation}
\lim_{\stackrel{N,M\to\infty}{M/N=\alpha}}\langle g(\lambda)\rangle_{A,\bm\xi}=\frac{1}{(\sqrt{s_+}-\sqrt{s_-})^2}\frac{(\sqrt{s_+ -\lambda}-\sqrt{s_- -\lambda})^2}{\sqrt{(s_+-\lambda)(s_- -\lambda)}}\ ,\label{doubleaverageG}
\end{equation}
for $\lambda<s_-$. The value $\lambda^\star$ will therefore be obtained from Eq. \eqref{equatingglambda} by equating \eqref{doubleaverageG} to $1/\sigma^2$, which leads after simple algebra to Eq. \eqref{lambdastar}. 

Note from Eq. \eqref{lambdastar} that:
\begin{enumerate}
\item For $\sigma^2\to 0$, we have $\lambda^\star\to (\sqrt{\alpha}-1)^2\equiv s_-$. Indeed, in this limit (see Fig. \ref{figG}), the smallest Lagrange multiplier should coincide with the minimal eigenvalue of the Wishart matrix.
\item From Fig. \ref{figG}, one may see that by increasing $\sigma$ for fixed $\alpha, N$ one makes the lowest Lagrange multiplier change sign from positive to negative.
Our calculation confirms this intuition and shows that this sign change typically happens when $\alpha = 1+\sigma^2$.
\end{enumerate}

\subsection{Calculation of $\langle  {\cal E}_{min}\rangle_{A,\bm b}$}
	
Having computed the typical value $\lambda^\star$ of the minimal solution of Eq. \eqref{equatingglambda} (averaged over the disorder) we insert this fixed, non-random value into Eq. \eqref{defEmin2} and proceed to compute the average of the loss function as follows:
\begin{align}
\nonumber &\langle  {\cal E}_{min}\rangle_{A,\bm b}=\frac{1}{2}\left\langle\Big|\Big|A(W-\lambda^\star\mathbb{I}_N)^{-1} A^T\bm b-\bm b\Big|\Big|^2\right\rangle_{A,\bm b}\\
\nonumber &=\frac{1}{2} \left\langle\left(A(W-\lambda^\star\mathbb{I}_N)^{-1} A^T\bm b-\bm b\right)^T \left(A(W-\lambda^\star\mathbb{I}_N)^{-1} A^T\bm b-\bm b\right)\right\rangle_{A,\bm b}\\
\nonumber &=\frac{1}{2} \left\langle\left(\bm b^T A(W-\lambda^\star\mathbb{I}_N)^{-1} A^T-\bm b^T\right) \left(A(W-\lambda^\star\mathbb{I}_N)^{-1} A^T\bm b-\bm b\right)\right\rangle_{A,\bm b}\\
\nonumber & =\frac{1}{2}\left\langle \bm b^T A(W-\lambda^\star\mathbb{I}_N)^{-1} \underbrace{A^T A}_W(W-\lambda^\star\mathbb{I}_N)^{-1} A^T\bm b -2 \bm b^T A (W-\lambda^\star\mathbb{I}_N)^{-1} A^T\bm b\right.\\
\nonumber &\hspace{30pt}\left.  +\bm b^T\bm b  \right\rangle_{A,\bm b}\ .
\end{align}
Replacing $W$ with $(W-\lambda^\star \mathbb{I}_N)+\lambda^\star \mathbb{I}_N$ in the middle of the first contribution leads to
\begin{align}
\nonumber &\langle  {\cal E}_{min}\rangle_{A,\bm b}=\frac{1}{2}\left\langle \bm b^T A(W-\lambda^\star\mathbb{I}_N)^{-1} A^T\bm b+\lambda^\star  \underbrace{\bm b^T A(W-\lambda^\star\mathbb{I}_N)^{-2} A^T\bm b}_{=N}\right.\\
&\left. -2 \bm b^T A (W-\lambda^\star\mathbb{I}_N)^{-1} A^T\bm b + \bm b^T\bm b \right\rangle_{A,\bm b}\ ,
\end{align}
where we used Eq. \eqref{sphconstraint}. Combining the first and third contribution together, we eventually get to
\begin{equation}
\langle  {\cal E}_{min}\rangle_{A,\bm b}=\frac{1}{2}\left\langle \bm b^T A(\lambda^\star\mathbb{I}_N-W)^{-1} A^T\bm b+\lambda^\star N+|\bm b|^2\right\rangle_{A,\bm b}\ .\label{finalEmin}
\end{equation}
	
We first perform the average over the i.i.d. Gaussian distributed $\bm b$ vector with variance $\sigma^2$, using
\begin{align}
\langle |\bm b|^2\rangle_{\bm b} &=\sum_{i=1}^M \langle b_i^2\rangle_{\bm b}=M\sigma^2\label{id1}\\ 
\langle \bm b^T R\bm b\rangle_{\bm b} &=\sum_{i,j} R_{ij} \langle b_i b_j\rangle_{\bm b}=\sigma^2 \sum_{i,j} R_{ij}\delta_{ij}=\sigma^2 \mathrm{Tr} R\label{id2}\ ,
\end{align}	
where $\mathrm{Tr}$ denotes the matrix trace.

Setting $R=A(\lambda^\star\mathbb{I}_N-W)^{-1} A^T$ and looking at the limit $N,M\to\infty$ with $\alpha=M/N$ fixed we get from Eq. \eqref{finalEmin}
\begin{equation}
\lim_{N,M\to\infty} \frac{\langle  {\cal E}_{min}\rangle_{A,\bm b}}{N}=\frac{1}{2}\left[\lambda^\star+\alpha\sigma^2+\sigma^2 \left\langle\frac{1}{N}\mathrm{Tr}\frac{W}{\lambda^\star\mathbb{I}_N-W}\right\rangle_A\right]\ .\label{almostfinalE}
\end{equation}	
The average over $A$ of the last term can be performed in the large $N,M$ limit appealing again to the formula \eqref{traceproperty} to yield
\begin{equation}
 \left\langle\frac{1}{N}\mathrm{Tr}\frac{W}{\lambda^\star\mathbb{I}_N-W}\right\rangle_A= \int_{s_-}^{s_+}\mathrm{d}s~\rho_{MP}(s)\frac{s}{\lambda^\star - s}\ ,
\end{equation}
which can be computed along the same lines as in Appendix \ref{app:integralMP} to give
\begin{align}
 \left\langle\frac{1}{N}\mathrm{Tr}\frac{W}{\lambda^\star\mathbb{I}_N-W}\right\rangle_A &=\frac{2}{\pi}\frac{(s_+ - s_-)^2}{(\sqrt{s_+}-\sqrt{s_-})^2 (\lambda-s_-)}\int_{0}^{1}\mathrm{d}t~\frac{\sqrt{t(1-t)}}{1+\gamma t}\\
 &= -\frac{\left(\sqrt{s_-}+\sqrt{s_+}\right)^2}{-2 \lambda^\star +2 \sqrt{(s_- -\lambda^\star) (s_+ -\lambda )}+s_- +s_+}\ ,
\end{align}
where 
\begin{equation}
\gamma = \frac{s_+ - s_-}{s_- -\lambda^\star}>0\ .
\end{equation}
Inserting this result back into Eq. \eqref{almostfinalE}, and substituting the value of $\lambda^\star$ as given in Eq. \eqref{lambdastar} as well as $s_\pm=(\sqrt{\alpha}\pm 1)^2$, we obtain after simplifications
\begin{equation}
\boxed{\lim_{N,M\to\infty} \frac{\langle  {\cal E}_{min}\rangle_{A,\bm b}}{N}=\frac{1}{2}\left[\sqrt{\alpha (1+\sigma^2)}-1\right]^2}\ .\label{finalEminLagrange}
\end{equation}
This result, obtained under the assumption\footnote{The assumption $\alpha>1$ (equivalently, $M>N$) was at play around Eq. \eqref{grouping} and \eqref{der2} (in the case $\alpha<1$ the matrix $W$ of size $N\times N$ would effectively be an anti-Wishart matrix, with $M<N$ nonzero eigenvalues). Also the evaluation of the integral \eqref{traceproperty} would require the spectral density of the anti-Wishart ensemble, given in Eq. \eqref{rhoAW}. It is likely that the RMT treatment of the case $\alpha<1$ could be carried out in full as well, however this is left for future work. } $\alpha>1$, confirms the natural expectation that over-complete systems with an additional quadratic constraint are typically incompatible, as the average cost function is strictly positive. However, the expression in \eqref{finalEminLagrange} vanishes for 
\begin{equation}
\alpha =\alpha_c=\frac{1}{1+\sigma^2}<1\ ,\label{alphacRMT}
\end{equation}
so -- assuming that the result in \eqref{finalEminLagrange} remains valid for $\alpha<1$ as well -- we may conclude that there are \emph{three} regimes for this problem:
\begin{itemize}
\item Over-complete systems ($\alpha>1$), which are typically incompatible. Here, the spherical constraint does not add much to what would happen anyway for ``standard'' linear systems (see Appendix \ref{appnospherical} for the calculation of the average minimal cost for linear systems without the spherical constraint).
\item Strongly under-complete systems ($0<\alpha<\alpha_c<1$), which are typically compatible even with the additional spherical constraint. 
\item Weakly under-complete systems  ($\alpha_c<\alpha<1$), which are typically incompatible. This is a new regime, where the additional spherical constraint is responsible for turning a linear system that would be perfectly compatible into an incompatible one due to the additional quadratic requirement to be satisfied by the unknowns.
\end{itemize}
It is interesting to notice that the critical threshold $\alpha_c$ shrinks with increasing variance $\sigma^2$ of the ``noise'' variables $b_k$. This means that the incompatibility region widens, and even systems with a handful of equations and many more unknowns become hard to satisfy as the known coefficients become more and more ``blurred''.

\section{Replica Calculation}\label{sec:replicas}

 To verify the formula \eqref{finalEminLagrange} for the minimal cost independently (and in this way also to justify some assumptions used to arrive to it), we now follow \cite{FLD2013} and reformulate the minimisation problem from the viewpoint of Statistical Mechanics of disordered systems.
 Namely, to look for the minimum of a loss/cost function $H (\bm x)$ over the sphere $||{\bm x }||^2=\textcolor{ForestGreen}{N}$ one
introduces an auxiliary non-negative parameter $\textcolor{red}{\beta}>0$ (called {\it the inverse temperature}) and uses it to define the so-called {\it canonical partition function} $Z(\textcolor{red}{\beta} )$ via
	\begin{equation}\label{method Z_intro}
		Z(\textcolor{red}{\beta} ) = \int\limits_{||{\bm x }||^2=\textcolor{ForestGreen}{N}} \mathrm{d} \bm x \, \mathrm{e}^{-\textcolor{red}{\beta}  H(\bm x)}\ .
	\end{equation}
	Suppose the minimum $\mathcal E_{min}$ of the loss/cost function $H (\bm x)$ is achieved at some $\bm x_{min}$. For $\textcolor{red}{\beta}\to\infty$, the integral in Eq. \eqref{method Z_intro} can be evaluated using Laplace's method (summarised in Appendix \ref{appLaplace}). Applying this method gives the leading exponential behaviour in the form 
	\begin{equation*}
		Z(\textcolor{red}{\beta}) \propto \mathrm{e}^{-\textcolor{red}{\beta}  \mathcal E_{min}}\ ,
	\end{equation*}
hence the minimal value can be recovered in the limit $\textcolor{red}{\beta} \to \infty$ from the logarithm of the partition function:
	\begin{equation}\label{min_free_energy_intro}
		 \mathcal E_{min}  = -\lim_{\textcolor{red}{\beta} \to\infty}\frac{1}{\textcolor{red}{\beta} }  \log Z(\textcolor{red}{\beta}) \ .
	\end{equation}
The canonical partition function is nothing but the normalisation factor of the Gibbs-Boltzmann probability density function
\begin{equation}
P_{\textcolor{red}{\beta}}(\bm x)=\frac{\mathrm{e}^{-\textcolor{red}{\beta}  H(\bm x)}}{Z(\textcolor{red}{\beta} ) }
\end{equation}
governing the probability of finding a thermodynamical system of particles around the point $\bm x$ in space, if their total energy is given by $H(\bm x)$ and they are in equilibrium with a heat bath at inverse temperature $\textcolor{red}{\beta}$. In this language, the \emph{free energy} of the system is given by
\begin{equation}
F(\textcolor{red}{\beta})= -\frac{1}{\textcolor{red}{\beta} }  \log Z(\textcolor{red}{\beta})\ ,
\end{equation}
from which it follows that $\mathcal E_{min}$ is nothing but the free energy at zero temperature of the associated particle system, and $\bm x_{min}$ the most likely zero-temperature configuration of particles at equilibrium.

 The relation \eqref{min_free_energy_intro} is very general (in particular, it does not assume the global minimum is achieved at only a single $\bm x_{min}$, there may be several such points), but for problems involving random cost/loss functions its actual usefulness crucially depends on our ability to characterise the behaviour of the log on the right-hand side.

 \subsection{Average $\langle  \mathcal E_{min}\rangle_{A,\bm b}$}\label{sec:replicas2}

 The simplest, yet already highly nontrivial task is to find the mean value  $\langle \mathcal E_{min} \rangle_{A,\bm b}$ for the minimal cost. Obviously, one needs to average the logarithm $ \log Z(\textcolor{red}{\beta})$ on the right-hand side of \eqref{min_free_energy_intro}
  \begin{equation}
 \langle \log Z(\textcolor{red}{\beta})  \rangle_{A,\bm b}=?\label{annoying}
 \end{equation}
 
 What is the problem here? Writing the average more explicitly, we are after the multiple integral
 \begin{equation}
 \int \mathrm{d}A~\mathrm{d}\bm b~ P_A(A)P_{\bm b}(\bm b)\log\left(\int\limits_{||{\bm x }||^2=\textcolor{ForestGreen}{N}} \mathrm{d} \bm x \, \mathrm{e}^{-\textcolor{red}{\beta}  H(\bm x)}\right)\ ,\label{quenched}
 \end{equation}
 where $P_A(A)$ and $P_{\bm b}(\bm b)$ are the joint probability density function of the entries of the matrix $A$, and of the vector $\bm b$, respectively\footnote{In our setting, these are just given by product of independent Gaussian functions, one for each entry.}. 
 Eq. \eqref{quenched} defines a so called \emph{quenched}\footnote{From Merriam-Webster dictionary. \emph{Quench} [transitive verb]: to cool (something, such as heated metal) suddenly by immersion (as in oil or water). } average: there are two nested sources of randomness, (i) the disorder in $A$ and $\bm b$, and (ii) the Gibbs-Boltzmann distribution at inverse temperature $\textcolor{red}{\beta}$ of the auxiliary degrees of freedom $\bm x$. A single instance of $A$ and of $\bm b$ is selected and kept fixed, while the $\bm x$ degrees of freedom equilibrate at inverse temperature $\textcolor{red}{\beta}$, before another instance of the $A,\bm b$ disorder is picked, and the process is repeated. In thermodynamical terms, a quenched average rests on the separation between two temporal scales, the ``fast'' equilibration of the $\bm x$ degrees of freedom at fixed temperature and at fixed instance of the disorder, and a ``slower'' scale of change of the background disorder. This should be contrasted with the \emph{annealed}\footnote{\emph{Anneal} [transitive verb]: to heat and then cool slowly (a material, such as steel or glass) usually for softening and making less brittle.} scheme, which is only approximate but way easier to handle analytically, consisting in treating the averages over the disorder and over the Gibbs-Boltzmann distribution on the same footing, and simultaneously. We will not consider annealed calculations here -- for more information, see \cite{CCrev,zamponi}.

 Clearly, the only obvious way to crack the integral \eqref{quenched} is to solve the $\bm x$-integral \emph{first}, take the logarithm of the result, and take the $A$ and $\bm b$ integrals \emph{last}. Unfortunately, this procedure fails in most cases, because the $\bm x$-integral cannot in general be solved exactly for a fixed instance of $A$ and $\bm b$ (or -- for different models -- of the so called \emph{disorder}, namely the randomness inherent to the parameters entering the definition of the cost function $H(\bm x)$ -- in our case, the entries of $A$ and $\bm b$).
 
 It would therefore be very helpful to make some headway if we were able somehow to take the average over the disorder ($A$ and $\bm b$) \emph{first}, and the $\bm x$-integral \emph{last} -- in the hope that swapping the order of integration would make the integrals more friendly to tackle.
 
 A heuristic recipe to do just that originated in the ``theory of spin glasses'' \cite{CCrev} in the 70s.  Doing this swapping of integrals in a fully rigorously manner is quite challenging even in the simplest possible instances, though considerable progress has been achieved in the last decades in evaluating such averages in a mathematically controllable way in the case when the cost function is normally distributed,  see e.g.~\cite{TalSpinGlass,bovier,panchenko}.  Unfortunately, in our case the cost function is not normally distributed, but rather represents a sum of squared normally distributed pieces. In such a case the rigorous theory has not been yet developed, but progress is still possible within the powerful but heuristic method of Theoretical Physics, known as the ``replica trick'', see e.g. \cite{PUZ2020}.  
 
 The main idea rests on the exact identity
 \begin{equation}
 \langle \log z \rangle  = \lim_{n\to 0} \frac{1}{n} \log \langle z^n\rangle\ ,\label{logidentity}
 \end{equation}
 which can be proven by noting that $\langle z^n\rangle=\langle 1 + n\log z+o(n)\rangle=\langle 1\rangle + n\langle \log z\rangle + \ldots =1 + n \langle\log z\rangle+\ldots$, where we used linearity of expectations, and normalisation $\langle 1\rangle=1$.
 
 While the identity \eqref{logidentity} is mathematically fully rigorous, and requires $n$ to be real and in the vicinity of zero, the way it is implemented in replica calculations is as follows: assume first that $\textcolor{blue}{n}$ is an integer (we will highlight this by colouring $\textcolor{blue}{n}$ in blue hereafter).

 This approach assumes  that the mean value we are after can be found not from directly calculating the average, but by considering the expectation of the integer moments of the partition function, frequently called in the physical literature the ``replicated'' disorder averaged partition function $\langle Z^{\textcolor{blue}{n}}(\textcolor{red}{\beta})  \rangle$  and subsequently taking the limit\footnote{We use the wiggly arrow $\rightsquigarrow$ to denote the ``replica limit'' procedure as follows: first convert $n\in\mathbb{R}$ to an integer. Then evaluate the corresponding observable as an explicit function of the integer $\textcolor{blue}{n}$. Then pretend that $\textcolor{blue}{n}$ could be analytically continued in the vicinity of zero without ambiguities.} $\textcolor{blue}{n}\rightsquigarrow 0$ to recover the averaged log
	\begin{equation}\label{replica method_intro}
		\langle  \mathcal E_{min} \rangle_{a,\bm b} = \lim_{\textcolor{red}{\beta} \to\infty}\frac{-1}{\textcolor{red}{\beta} }\lim\limits_{\textcolor{blue}{n}\rightsquigarrow 0} \frac{1}{\textcolor{blue}{n}} \log{\langle Z^{\textcolor{blue}{n}}(\textcolor{red}{\beta}) \rangle}_{A,\bm b}
	\end{equation}	
and eventually for large $N,M$ with $\alpha=M/N$ constant, we wish to compute
	\begin{equation}\label{replica method_intro2}
	\lim_{\textcolor{ForestGreen}{N}\to\infty}	\frac{\langle  \mathcal E_{min} \rangle_{A,\bm b}}{\textcolor{ForestGreen}{N}}= 	\lim_{\textcolor{ForestGreen}{N}\to\infty}	\frac{1}{\textcolor{ForestGreen}{N}} \lim_{\textcolor{red}{\beta} \to\infty}\frac{-1}{\textcolor{red}{\beta} }\lim\limits_{\textcolor{blue}{n}\rightsquigarrow 0} \frac{1}{\textcolor{blue}{n}} \log{\langle Z^{\textcolor{blue}{n}}(\textcolor{red}{\beta}) \rangle}_{A,\bm b}
	\end{equation}
to be compared with the corresponding RMT result in Eq. \eqref{finalEminLagrange}.

We now show how to evaluate the replicated moments of the finite-temperature partition function induced by a general least-square random cost landscape of the form
\begin{equation}\label{cost_gen_poly}
		H (\bm x) = \frac{1}{2} \sum\limits_{k=1}^M \left[V_k(\bm x)\right]^2\ ,
	\end{equation}
restricted to the sphere $||{\bm x}||^2=\textcolor{ForestGreen}{N}$, where $V_k(\bm{x})$ are chosen to be Gaussian-distributed functions of the vector $\bm x$ with expectations $\langle V_k(\bm x)\rangle_V=0$ and covariance
\begin{equation}\label{cov_gen}
\langle V_k(\bm x_a) V_\ell (\bm x_b)\rangle_V = \delta_{k\ell} f\! \left(\frac{\bm x_a \cdot \bm x_b}{\textcolor{ForestGreen}{N}} \right)\ .
\end{equation}
Here and in the following, $\bm u\cdot \bm v$ denotes the dot product between vectors.

The particular case $V_k(\bm x)=\sum_{j=1}^{\textcolor{ForestGreen}{N}} A_{kj}x_j-b_k$ treated here (linear system of equations) corresponds to the choice 

\begin{equation}
f(u)=\sigma^2+u\label{deffu}
\end{equation}
(see Appendix \ref{app:covariance} for the full derivation), but the calculation can be performed for any covariance of the form \eqref{cov_gen}
and essentially follows the method of  \cite{Fyo2019}. The non-linear version of the model was considered in \cite{tublin} and then fully solved in \cite{urbanirig} in the context of a model of rigidity transition in confluent systems, where a full Replica Symmetry Breaking structure was uncovered and the properties around the transition were studied in detail (see also \cite{urb1} for an insightful dynamical mean-field theory of the non-linear model, and \cite{urb2} for a quantum version of it).

Starting with the partition function $Z(\textcolor{red}{\beta})$ 
\begin{equation}\label{method Z_Appendix1}
		Z(\textcolor{red}{\beta}) = \int\limits_{||{\bm x}||^2=\textcolor{ForestGreen}{N}} \mathrm{d} \bm x~\mathrm{e}^{-\frac 12 \textcolor{red}{\beta}   \sum\limits_{k=1}^M V_k^2 (\bm x)}\ ,
\end{equation}
we aim to arrive to a convenient integral representation for the averaged partition function of $\textcolor{blue}{n}$ copies of the same system $\langle Z^{\textcolor{blue}{n}} (\textcolor{red}{\beta})\rangle$.	To facilitate the averaging, we introduce new auxiliary variables of integration and employ the standard Gaussian integral identity (sometimes called in physics literature the Hubbard-Stratonovich identity) 	 
\begin{equation}
\mathrm{e}^{-\frac{\textcolor{red}{\beta}}{2}\xi^2}=\frac{1}{\sqrt{2\pi}}\int_{\mathbb{R}}\mathrm{d}u~\exp\left[-\frac{1}{2}u^2-\mathrm{i}\sqrt{\textcolor{red}{\beta}}\xi u\right]\ ,
\end{equation}	
applying it $M$ times, once for every index $k=1,\ldots,M$. The effect of this identity is to lower the power of the quantity $\xi$ (in our case, each $\xi_k$ is equal to $V_k(\bm x)$), from $2$ to $1$ -- the price to pay is that one needs to introduce an extra integration over an auxiliary Gaussian variable $u$.

Proceeding in this way the partition function $Z(\textcolor{red}{\beta})$ now can be expressed in terms of the integration over a vector ${\bm u = (u_1, \dots , u_M)^T}$ as
\begin{equation}
Z(\textcolor{red}{\beta}) = \int\limits_{\mathbb R^M} \frac{\mathrm{d} \bm u}{ (2\pi)^{M/2}}~\mathrm{e}^{-\frac 12 (\bm u \cdot  \bm u)} \int\limits_{||{\bm x}||^2=\textcolor{ForestGreen}{N}} \mathrm{d} \bm x~\mathrm{e}^{-\mathrm{i} \sqrt{\textcolor{red}{\beta} } \sum\limits_{k=1}^M u_k V_k (\bm x)}\ .
\end{equation}

 Now we take products of \textcolor{blue}{$n$} identical copies of $Z(\textcolor{red}{\beta})$ (which we number with the index $a=1,2,\ldots, \textcolor{blue}{n}$) and aim to evaluate 
	\begin{align}
		\nonumber\langle Z^{\textcolor{blue}{n}}(\textcolor{red}{\beta}) \rangle_V &=  \int\limits_{\mathbb R^{\textcolor{blue}{n}M}} \prod_{a=1}^{\textcolor{blue}{n}} \frac{\mathrm{d} \bm u_a}{ (2\pi)^{M/2}} \mathrm{e}^{-\frac{1}{2} \sum\limits_{a=1}^{\textcolor{blue}{n}} (\bm u_a \cdot \bm u_a)} \times\\
		&\times \int_{D_{\textcolor{ForestGreen}{N}}} \prod\limits_{a=1}^{\textcolor{blue}{n}} \mathrm{d} \bm x_a \prod\limits_{k=1}^M\left\langle \mathrm{e}^{-\mathrm{i} \sqrt{\textcolor{red}{\beta} }  \sum\limits_{a=1}^{\textcolor{blue}{n}} [\bm u_a]_k V_k (\bm x_a)} \right\rangle_V\ ,\label{averageV}
	\end{align}
	where rectangular brackets $[\bm u_a]_k$ stand for the $k$-th entry of the vector $\bm u_a$, and $\langle\cdot\rangle_V$ stands for averaging over the random variable $V$. Here we used the fact that
for different $k$ the functions $V_k(\bm x)$ are independent, hence the corresponding average of the product factorises in the product of averages. The domain of integration of the $\bm x$ variables is the union of $\textcolor{blue}{n}$ spheres $D_{\textcolor{ForestGreen}{N}} = \{  \, ||{\bm x_a}||^2 = \textcolor{ForestGreen}{N}, \,  \forall a \}$.

Note that replicating the partition function an \emph{integer} number of times $\textcolor{blue}{n}$ has effectively allowed us to get rid of the annoying logarithm in \eqref{annoying} and offers now a way to perform the average over the randomness in $A$ and $\bm b$ (through $V$) directly.

	 To perform the average over $V_k$  we use that the combination\\
$z=-\mathrm{i} \sqrt{\textcolor{red}{\beta} }  \sum\limits_{a=1}^{\textcolor{blue}{n}} [\bm u_a]_k V_k (\bm x_a)$ is Gaussian -- being a sum of Gaussian-distributed terms $V_k$ -- with mean zero and variance
\begin{equation}
\langle z^2\rangle_V=-\textcolor{red}{\beta} \sum\limits_{a,b=1}^{\textcolor{blue}{n}} [\bm u_a]_k[\bm u_b]_k \langle V_k (\bm x_a)V_k (\bm x_b)\rangle_V =-\textcolor{red}{\beta} \sum\limits_{a,b=1}^{\textcolor{blue}{n}} [\bm u_a]_k[\bm u_b]_k     f\! \left(\frac{\bm x_a \cdot \bm x_b}{\textcolor{ForestGreen}{N}} \right)\ ,
\end{equation}
where we have used the covariance of the $V$'s in Eq.  \eqref{cov_gen}. 

Using now the following rule for computing the expectation of the exponential of a Gaussian variable
\begin{equation}
\langle \mathrm{e}^z\rangle_{z\sim\mathcal{N}(\mu,\sigma^2)}=\mathrm{e}^{\mu+\frac{\sigma^2}{2}}
\end{equation}
and taking the product over $k=1,\ldots,M$, we arrive at
	\begin{align}
\nonumber		\prod\limits_{k=1}^M &\left\langle \mathrm{e}^{-\mathrm{i} \sqrt{\textcolor{red}{\beta} } \sum\limits_{a=1}^{\textcolor{blue}{n}} [\bm u_a]_k V_k (\bm x_a)} \right\rangle_V
		= \prod\limits_{k=1}^M \exp\left[-\frac 12 \textcolor{red}{\beta}  \sum_{a,b=1}^{\textcolor{blue}{n}} [\bm u_a]_k [\bm u_b]_k f\! \left(\frac{\bm x_a \cdot \bm x_b}{\textcolor{ForestGreen}{N}} \right) \right]\\
	\nonumber	&=\exp\left[-\frac 12 \textcolor{red}{\beta}  \sum_{a,b=1}^{\textcolor{blue}{n}} \left(\sum_{k=1}^M[\bm u_a]_k [\bm u_b]_k \right)f\! \left(\frac{\bm x_a \cdot \bm x_b}{\textcolor{ForestGreen}{N}} \right) \right]\\
		&= \exp\left[-\frac 12 \textcolor{red}{\beta}  (\bm u_a \cdot \bm u_b) f\! \left(\frac{\bm x_a \cdot \bm x_b}{\textcolor{ForestGreen}{N}} \right) \right]\ .\label{averageV2}
	\end{align}
Substituting this back into the integral representation for $\langle Z^{\textcolor{blue}{n}}(\textcolor{red}{\beta})\rangle_V$ in Eq. \eqref{averageV}, and writing the dot products more explicitly we get

\begin{align}
		\nonumber\langle Z^{\textcolor{blue}{n}}(\textcolor{red}{\beta}) \rangle_V &=\int_{D_{\textcolor{ForestGreen}{N}}} \prod\limits_{a=1}^{\textcolor{blue}{n}} \mathrm{d} \bm x_a  \int\limits_{\mathbb R^{\textcolor{blue}{n}M}} \prod_{a=1}^{\textcolor{blue}{n}} \frac{\mathrm{d} \bm u_a}{ (2\pi)^{M/2}} \exp\left[-\frac{1}{2} \sum_{k=1}^M\sum\limits_{a=1}^{\textcolor{blue}{n}} ([\bm u_a]_k)^2\right.\\
		&\left. -\frac{1}{2}\textcolor{red}{\beta} \sum_{k=1}^M  \sum_{a,b=1}^{\textcolor{blue}{n}} \left([\bm u_a]_k [\bm u_b]_k \right)f\! \left(\frac{\bm x_a \cdot \bm x_b}{\textcolor{ForestGreen}{N}} \right)     \right]\ .
	\end{align}
	
One notices by inspection that the exponential factorises into the product of $M$ identical exponentials, therefore we can write

\begin{align}
		\nonumber\langle Z^{\textcolor{blue}{n}}(\textcolor{red}{\beta}) \rangle_V &=\int_{D_{\textcolor{ForestGreen}{N}}} \prod\limits_{a=1}^{\textcolor{blue}{n}} \mathrm{d} \bm x_a  \left\{\int\limits_{\mathbb R^{\textcolor{blue}{n}}} \prod_{a=1}^{\textcolor{blue}{n}} \frac{\mathrm{d} u_a}{ (2\pi)^{M/2}} \exp\left[-\frac{1}{2} \sum\limits_{a=1}^{\textcolor{blue}{n}} u_a^2\right.\right.\\
		&\left.\left. -\frac{1}{2}\textcolor{red}{\beta}  \sum_{a,b=1}^{\textcolor{blue}{n}} u_a f\! \left(\frac{\bm x_a \cdot \bm x_b}{\textcolor{ForestGreen}{N}} \right)u_b     \right]\right\}^M\ .
	\end{align}
	
We notice now that the integral over $\{u_a\}$  is simply a multivariate $n$-dimensional Gaussian integration\footnote{The formula reads $\int_{\mathbb{R}^n}\mathrm{d}\bm u~\mathrm{e}^{-\frac{1}{2}\bm u^T M\bm u} =\sqrt{\frac{(2\pi)^n}{\det M}}$, for $M$ a $n\times n$ symmetric and positive definite matrix.}, yielding
	\begin{align}
		\langle Z^{\textcolor{blue}{n}}(\textcolor{red}{\beta})\rangle_V
		&\propto \int_{D_{\textcolor{ForestGreen}{N}}} \prod\limits_{a=1}^{\textcolor{blue}{n}} \mathrm{d} \bm x_a \left(\det \left[\mathbb{I}_{\textcolor{blue}{n}} + \textcolor{red}{\beta}  f \left(\frac{ \bm x_a \cdot \bm x_b}{N} \right) \right] \right)^{-\frac M2} \notag\\
		&\propto \int\limits_{D_{\textcolor{ForestGreen}{N}}^{(Q)}} \mathrm{d} Q (\det Q)^{\frac{\textcolor{ForestGreen}{N}-\textcolor{blue}{n}-1}{2}} \left( \det \left[\mathbb{I}_{\textcolor{blue}{n}} + \textcolor{red}{\beta}  \hat f (Q ) \right] \right)^{-\frac{\alpha \textcolor{ForestGreen}{N}}{2}} \ .\label{<Z^n> pre}
	\end{align}
	Here we omitted the exact proportionality constants (which are known but redundant for our goals), and we have replaced $M$ with $\alpha \textcolor{ForestGreen}{N}$ in view of a large $ \textcolor{ForestGreen}{N}$ asymptotics.
	
	 The change of variables in the last line  from the set of vectors ${\bm x_a}$ to the positive (semi)definite matrix $Q$ of size $\textcolor{blue}{n}\times \textcolor{blue}{n}$ with entries defined as\footnote{The values $q_{ab}$ are known in the spin-glass literature as the \emph{overlaps} between replicas, see \cite{CCrev} for details.} 
	 \begin{equation}
	 q_{ab}=\frac 1N (\bm x_a \cdot \bm x_b)\label{defqab}
	 \end{equation}
	  follows the idea of the paper \cite{Fyo2002}. The details of this transformations, including evaluation of the involved Jacobian determinant factor appearing in the above are explained in detail in ~\cite{FyoPhysA2010}, Eq. (47). The hat over $\hat f (Q)$ serves as a reminder that this is an $\textcolor{blue}{n}\times \textcolor{blue}{n}$ matrix as well, with entries $f_{ab}:=f(q_{ab})$. 	
	The domain of integration is then over such positive semi-definite matrices $Q$ with the constraint $q_{aa}=1$ on diagonal entries coming from \eqref{defqab} and the spherical constraint, that is given by
\begin{equation}
		Q = \begin{pmatrix}
			1 & & q_{ab} \\
			& \ddots & \\
			q_{ab} & & 1
		\end{pmatrix}.
	\end{equation}
In short, the integration goes over the domain   $D_{\textcolor{ForestGreen}{N}}^{(Q)} = {\{ Q\geq 0, q_{aa}=1 \, \forall a\}}$, and the integration measure is $\mathrm{d}Q\equiv\prod_{a<b}\mathrm{d}q_{ab}$.  
	  
Up to this step, no approximation has been used. Unfortunately, the matrix integral in Eq. \eqref{<Z^n> pre} is not easy to compute in closed form for any $\textcolor{ForestGreen}{N}$ -- result that would be needed to compute the various limits in the right hand side of Eq. \eqref{replica method_intro2}, in the exact order in which they appear (first \textcolor{blue}{$n$}, then \textcolor{red}{$\beta$}, and only at the end \textcolor{ForestGreen}{$N$}). Therefore, we are forced to walk another non-rigorous path, by changing the order of limits, and performing the $\textcolor{ForestGreen}{N}\to\infty$ limit \emph{first}.
	  
To this end, we rewrite the integral $\langle Z^{\textcolor{blue}{n}}(\textcolor{red}{\beta})\rangle_V$ in the form convenient for approximating it in the limit $\textcolor{ForestGreen}{N}\gg 1$ by Laplace's method:
	\begin{align}
		\langle Z^{\textcolor{blue}{n}}(\textcolor{red}{\beta})\rangle_V &\propto \int\limits_{D_{\textcolor{ForestGreen}{N}}^{(Q)}} \mathrm{d} Q \, (\det Q)^{-\frac{\textcolor{blue}{n}+1}{2}} \mathrm{e}^{  -\frac{\textcolor{ForestGreen}{N}}{2} \Phi_{\textcolor{blue}{n},\textcolor{red}{\beta}} (Q) } ,\label{intQ}\\
		& \Phi_{\textcolor{blue}{n},\textcolor{red}{\beta}} (Q)  =\alpha \log \det [\mathbb{I}_{\textcolor{blue}{n}} + \textcolor{red}{\beta}  \hat f (Q )  ] - \log \det Q\ ,
	\end{align}
	which we shall analyse for the particular choice $f(u)=\sigma^2+u$, corresponding to our problem (see Eq. \eqref{deffu}). In our setting, the function $\Phi_{\textcolor{blue}{n},\textcolor{red}{\beta}} (Q) $ therefore specialises to
\begin{equation}\label{special case}
\Phi_{\textcolor{blue}{n},\textcolor{red}{\beta}} (Q)  \equiv \alpha \log \det [ \mathbb{I}_{\textcolor{blue}{n}} + \textcolor{red}{\beta}  (Q+\sigma^2 E_{\textcolor{blue}{n}})] - \log \det Q\ ,
\end{equation}
 where we recall that $\alpha=M/\textcolor{ForestGreen}{N}$ is the ratio between number of equations and number of unknowns of our system, $\sigma^2$ is the variance of the ``noise'' terms $b_k$ in the linear system, $E_{\textcolor{blue}{n}}$ stands for the $\textcolor{blue}{n}\times \textcolor{blue}{n}$ matrix with all entries equal to unity, whereas
 $ \mathbb{I}_{\textcolor{blue}{n}}$ stands for the $\textcolor{blue}{n}\times \textcolor{blue}{n}$ identity matrix.
 
Due to the presence of large factor $\textcolor{ForestGreen}{N}$ in the exponent under the integral over $Q$ in \eqref{intQ}, we can apply the Laplace's approximation, which then implies that for some extremising matrix argument $Q_{extr}$
\begin{equation}
\langle Z^{\textcolor{blue}{n}}(\textcolor{red}{\beta})\rangle_V \approx \mathrm{e}^{-\frac{\textcolor{ForestGreen}{N}}{2} \Phi_{\textcolor{blue}{n},\textcolor{red}{\beta}}  (Q_{extr})}\ ,\label{laplaceZQ}
\end{equation}
where we keep only the leading exponential term, as this is the only one needed to compute $ \langle  \mathcal E_{min} \rangle_{A,\bm b}$ to the leading order (see Eq. \eqref{replica method_intro2} after exchanging the order of limits there)
	\begin{align}
	\nonumber \lim_{\textcolor{ForestGreen}{N}\to\infty}	\frac{\langle  \mathcal E_{min} \rangle_{A,\bm b}}{\textcolor{ForestGreen}{N}} &= 	\lim_{\textcolor{ForestGreen}{N}\to\infty}	\frac{1}{\textcolor{ForestGreen}{N}} \lim_{\textcolor{red}{\beta} \to\infty}\frac{-1}{\textcolor{red}{\beta} }\lim_{\textcolor{blue}{n}\rightsquigarrow 0} \frac{1}{\textcolor{blue}{n}} \log{\langle Z^{\textcolor{blue}{n}}(\textcolor{red}{\beta}) \rangle}_{A,\bm b}\\
	&\simeq  \lim_{\textcolor{red}{\beta} \to\infty}\frac{-1}{\textcolor{red}{\beta} }\lim_{\textcolor{blue}{n}\rightsquigarrow 0} \frac{1}{\textcolor{blue}{n}} \lim_{\textcolor{ForestGreen}{N}\to\infty}	\frac{1}{\textcolor{ForestGreen}{N}} 
	\log{\langle Z^{\textcolor{blue}{n}}(\textcolor{red}{\beta}) \rangle}_{A,\bm b}\\
	&\simeq  \lim_{\textcolor{red}{\beta} \to\infty}\frac{-1}{\textcolor{red}{\beta} }\lim_{\textcolor{blue}{n}\rightsquigarrow 0} \frac{1}{\textcolor{blue}{n}} \lim_{\textcolor{ForestGreen}{N}\to\infty}	\frac{1}{\textcolor{ForestGreen}{N}} 
	\log  \mathrm{e}^{-\frac{\textcolor{ForestGreen}{N}}{2} \Phi_{\textcolor{blue}{n},\textcolor{red}{\beta}}  (Q_{extr})}\\
	&\simeq  \lim_{\textcolor{red}{\beta} \to\infty}\frac{1}{2\textcolor{red}{\beta} }\lim_{\textcolor{blue}{n}\rightsquigarrow 0} \frac{1}{\textcolor{blue}{n}} \Phi_{\textcolor{blue}{n},\textcolor{red}{\beta}}  (Q_{extr})\ ,\label{replica method_intro2v2}
	\end{align}
where we first exchanged the order of integration (non-rigorous step), and then we inserted the asymptotic behaviour \eqref{laplaceZQ} coming from the Laplace approximation.

We now have to determine $Q_{extr}$ using the stationarity conditions
\begin{equation}
\frac{\partial   \Phi_{\textcolor{blue}{n},\textcolor{red}{\beta}}  (Q)}{\partial q_{ab}}=0, \qquad\forall a<b\ ,
\end{equation}
where we restrict to the upper triangle of the matrix $Q$ since it is symmetric.

Looking at Eq. \eqref{special case}, we need to use the formula
\begin{equation}
\frac{\partial}{\partial M_{ab}}\log\det M = [M^{-1}]_{ba}\ ,\label{formulalogdet}
\end{equation}
valid for an invertible square matrix $M$. A proof is provided in Appendix \ref{app:identities}.

Applying this identity to Eq. \eqref{special case} and setting the result to zero, we obtain an equation for the $ab$ entry of two inverse matrices, both related to $Q_{extr}$
\begin{equation}\label{stationarity_simple}
\left(Q_{extr}^{-1}\right)_{ab}=\alpha \textcolor{red}{\beta}  \left[\left(\mathbb{I}_{\textcolor{blue}{n}} + \textcolor{red}{\beta}  (Q_{extr}+\sigma^2 E_{\textcolor{blue}{n}})\right)^{-1}\right]_{ab}\ .
\end{equation}
To solve this equation we use the so-called {\it Replica Symmetric} (RS) ansatz, which amounts to assuming that all diagonal entries of the matrix $Q_{extr}$ are equal to one and all other entries are the same number $q$. At first sight,
no other ansatz would seem to make any sense, as all replicas are ``created equal'' -- therefore, two different entries $q_{ab}=\bm x_a\cdot \bm x_b/N$ and $q_{cd}=\bm x_c\cdot \bm x_d/N$ of the matrix $Q_{extr}$ do not have any reason to behave differently from each other at the saddle point. Surprisingly, it was shown by Parisi \cite{Parisi79} in the context of the solution of the Sherrington-Kirkpatrick model of spin glasses that the RS scheme may actually fail to produce physical results in certain cases, and should be replaced by an elaborate hierarchy of more sophisticated ans\"atze where the seemingly obvious symmetry between replicas is broken at various levels. Fortunately, the RS ansatz is sufficient for our problem here -- why this is so, and what to do when it is not, goes beyond the scope of these lecture notes and will not be addressed here.

Therefore

\begin{equation}
Q_{extr} = 
\left( 
\begin{array}{ccccc}
1 & & & & \raisebox{-3ex}[0pt][0pt]{\Huge $q$} \\
 & 1 & & & \\
 & & \ddots & & \\
 & & & 1 & \\
\raisebox{3ex}[0pt][0pt]{\Huge $q$} & & & & 1 
\end{array}
\right)\ . \label{ansatzQ} 
\end{equation}


In order to ensure that the $\textcolor{blue}{n}\times \textcolor{blue}{n}$ matrix $Q_{extr}$ be positive semi-definite, we apply an extended version of the Sylvester's criterion \cite{SD} that states that a symmetric matrix is positive-semidefinite if and only if all its \emph{principal minors}\footnote{A principal minor of a matrix is the determinant of the sub-matrix obtained by erasing corresponding sets of rows and columns (e.g. rows $1$ and $6$, and columns $1$ and $6$).} are nonnegative. This results in the conditions
\begin{equation}
1-q\geq 0\qquad\text{\rm and}\qquad \frac{1}{2-\textcolor{blue}{n}}\leq q\leq 1\ .
\end{equation}
Given that -- at this stage -- the size of the matrix is an \emph{arbitrary} integer, it follows that we should restrict the value $q$ of the off-diagonal entries to the range $0\leq q\leq 1$.

We now make the following ansatz for its inverse

\begin{equation}
Q_{extr}^{-1} = 
\left( 
\begin{array}{ccccc}
\gamma & & & & \raisebox{-3ex}[0pt][0pt]{\Huge $\eta$} \\
 & \gamma & & & \\
 & & \ddots & & \\
 & & & \gamma & \\
\raisebox{3ex}[0pt][0pt]{\Huge $\eta$} & & & & \gamma 
\end{array}
\right)\ . \label{ansatz_inverse}
\end{equation}

It follows that
\begin{align}
[Q_{extr}Q_{extr}^{-1}]_{aa} &=\gamma+(\textcolor{blue}{n}-1)q\eta\\
[Q_{extr}Q_{extr}^{-1}]_{ab} &=q\gamma+\eta+(\textcolor{blue}{n}-2)q\eta\ ,
\end{align}
which should be set to $1$ and $0$, respectively. Solving for $\gamma$ and $\eta$, we find
\begin{align}
\gamma &= \frac{1+q(\textcolor{blue}{n}-2)}{(1-q)(1+q(\textcolor{blue}{n}-1))}\\
\eta &=\frac{-q}{(1-q)(1+q(\textcolor{blue}{n}-1))}\ .
\end{align}

Similarly, it is easy to find the inverse of the matrix $R=\mathbb{I}_{\textcolor{blue}{n}} + \textcolor{red}{\beta}  (Q_{extr}+\sigma^2 E_{\textcolor{blue}{n}})$ appearing on the right hand side of Eq. \eqref{stationarity_simple} since this matrix has
diagonal entries all equal to $R_{aa}=1+\textcolor{red}{\beta} (1+\sigma^2)=:r_d,\, \forall a=1,\ldots,\textcolor{blue}{n}$ and all off-diagonal entries  equal to $R_{ab}=\textcolor{red}{\beta} (q+\sigma^2)=:r, \, \forall a<b$. Using a similar matrix ansatz as in Eq. \eqref{ansatz_inverse}, we get

\begin{equation}\label{Rinverse}
\left(R^{-1}\right)_{aa}=\frac{\left(r_d+r(\textcolor{blue}{n}-2)\right)}{(r_d-r)\left(r_d+r(\textcolor{blue}{n}-1)\right)}, \quad \left(R^{-1}\right)_{a<b}=-\frac{r}{(r_d-r)\left(r_d+r(\textcolor{blue}{n}-1)\right)}\ .
\end{equation}

Hence the stationarity condition Eq. \eqref{stationarity_simple} for off-diagonal elements takes the form
\begin{equation}\label{stationarity_RS_simple}
\frac{q}{(1-q)\left(1+q(\textcolor{blue}{n}-1)\right)}=\alpha \frac{\textcolor{red}{\beta} ^2(q+\sigma^2)}{(1+\textcolor{red}{\beta} -\textcolor{red}{\beta}  q)(1+\textcolor{red}{\beta} -\textcolor{red}{\beta}  q+\textcolor{blue}{n}\textcolor{red}{\beta} (q+\sigma^2))}\ .
\end{equation}

We also have to compute $\Phi_{\textcolor{blue}{n},\textcolor{red}{\beta}} (Q_{extr}) $ from Eq. \eqref{special case}. To do so, we have to compute the determinant of matrices with the following structure
\begin{equation}
\det_{n\times n}\begin{pmatrix}
\gamma & \eta & \cdots & \eta\\
\vdots & \vdots & \ddots & \vdots\\
\eta & \eta & \cdots &\gamma
\end{pmatrix}=(\gamma-\eta)^{n-1}\left[\gamma+(n-1)\eta\right]\ ,
\end{equation}
a result that can be easily proven by induction.

Applying this formula to the two terms in Eq. \eqref{special case} after inserting the replica-symmetric ansatz Eq. \eqref{ansatzQ}, we have
 \begin{align}\label{Phi_RSa}
\nonumber\Phi_{\textcolor{blue}{n},\textcolor{red}{\beta}} (Q_{extr}) &=\alpha \textcolor{blue}{n}\log{\left(1+\textcolor{red}{\beta} (1-q)\right)}+\alpha\log{\left(1+\frac{\textcolor{red}{\beta} (q+\sigma^2)\textcolor{blue}{n}}{1+\textcolor{red}{\beta} (1-q)}\right)}\\
&-\textcolor{blue}{n}\log{(1-q)}-\log{\left(1+q\frac{\textcolor{blue}{n}}{1-q}\right)}\ ,
\end{align}
where $q$ is solution of the saddle-point equation \eqref{stationarity_RS_simple}.

So far, none of the steps infringed on the integer nature of the replica index $\textcolor{blue}{n}$. However, in Eq. \eqref{Phi_RSa}, $\textcolor{blue}{n}$ now appears as a \emph{parameter}, and no longer as the \emph{integer} number of integrals (as in Eq. \eqref{averageV}) or the \emph{integer} size of a matrix (as in Eq. \eqref{<Z^n> pre})\footnote{One of the most interesting aspects of replica calculations is indeed the fact that the replica parameter $\textcolor{blue}{n}$ changes nature and morphs into something else as the calculation proceeds.}. This is good news, as in the end we are seeking to perform the replica limit $\textcolor{blue}{n}\rightsquigarrow 0$. Treating now $\textcolor{blue}{n}$ as a real parameter that can get arbitrarily close to zero, we get
\begin{equation}\label{Phi_RSb}
\lim_{\textcolor{blue}{n}\rightsquigarrow 0} \frac{1}{\textcolor{blue}{n}} \Phi_{\textcolor{blue}{n},\textcolor{red}{\beta}}  (Q_{extr})=\alpha \log{\left(1+\textcolor{red}{\beta} (1-q)\right)}+\frac{\alpha\textcolor{red}{\beta} (q+\sigma^2)}{1+\textcolor{red}{\beta} (1-q)}-\log{(1-q)}-\frac{q}{1-q}\ ,
\end{equation}
where we used the standard asymptotics $\log(1+\epsilon)\sim \epsilon$ for small $\epsilon$.

We can also seek for the solution of the saddle-point equation \eqref{stationarity_RS_simple}
directly in the replica limit setting $\textcolor{blue}{n}=0$ in the equation, giving:
\begin{equation}\label{stationarity_RS_simple_replica}
\frac{q}{(1-q)^2}=\alpha\frac{\textcolor{red}{\beta} ^2(q+\sigma^2)}{(1+\textcolor{red}{\beta} -\textcolor{red}{\beta}  q)^2}\ ,
\end{equation}
to be solved for $q\in [0,1)$, where $q$ is the generic off-diagonal element of the extremising matrix $Q_{extr}$.

For finite $\textcolor{red}{\beta} $ the equation is equivalent to a cubic one. To simplify our considerations further, we recall that to find the average minimum of the cost function for large $\textcolor{ForestGreen}{N}$ from \eqref{replica method_intro2v2} we only need to know $q$ in the limit $\textcolor{red}{\beta}  \to \infty$. Two different situations may happen in this limit
\begin{enumerate}
\item  $q$ tends in the limit $\textcolor{red}{\beta} \to \infty$ to a non-negative value smaller than unity
(this range is dictated by positivity of the matrix $Q$).
\item  Alternatively, in such a limit $q$ tends to unity from below in such a way that
$v=\lim_{\textcolor{red}{\beta}  \to \infty} \textcolor{red}{\beta}  (1-q)$ remains finite.
 \end{enumerate}
 We can analyse the two situations separately.
 \begin{enumerate}
 \item Computing the limit $\textcolor{red}{\beta}  \to \infty$ on the right hand side and simplifying the denominator $(1-q)^2$, we obtain the equation
 \begin{equation}
 q=\alpha (q+\sigma^2)\Rightarrow q = \frac{\alpha\sigma^2}{1-\alpha}\ .
 \end{equation}
 Imposing the condition 
 \begin{equation}
 0\leq  \frac{\alpha\sigma^2}{1-\alpha}<1\Rightarrow \boxed{0<\alpha<\alpha_c:=\frac{1}{1+\sigma^2}<1}\ ,\label{alphacreplicas}
 \end{equation}
 where the same critical threshold $\alpha_c$ was obtained via a completely different route earlier on (see Eq. \eqref{alphacRMT}).
 
 Invoking Eq. \eqref{replica method_intro2v2} in the form
 \begin{align}
 \nonumber &\lim_{\textcolor{ForestGreen}{N}\to\infty}	\frac{\langle  \mathcal E_{min} \rangle_{A,\bm b}}{\textcolor{ForestGreen}{N}} =  \lim_{\textcolor{red}{\beta} \to\infty}\frac{1}{2\textcolor{red}{\beta} }\lim_{\textcolor{blue}{n}\rightsquigarrow 0} \frac{1}{\textcolor{blue}{n}} \Phi_{\textcolor{blue}{n},\textcolor{red}{\beta}}  (Q_{extr})\\
  &=  \lim_{\textcolor{red}{\beta} \to\infty}\frac{1}{2\textcolor{red}{\beta} }\left[\alpha \log{\left(1+\textcolor{red}{\beta} (1-q)\right)}+\frac{\alpha\textcolor{red}{\beta} (q+\sigma^2)}{1+\textcolor{red}{\beta} (1-q)}-\log{(1-q)}-\frac{q}{1-q}\right]=0\ ,
 \end{align}
 where we used the result in Eq. \eqref{Phi_RSb}, shows that the system of equations with a quadratic constraint is generally compatible if the number of equations is smaller than the number of unknowns (which is expected, and no different from the case with no quadratic constraint), but only up to a critical ratio $\alpha_c<1$. What happens beyond $\alpha_c$ is determined by the next case below. 
 \item We now seek for a solution of the saddle-point equation Eq. \eqref{stationarity_RS_simple_replica} for $\textcolor{red}{\beta}\to\infty$ and $q\to 1^-$, in such a way that the combination $\textcolor{red}{\beta}(1-q)=v>0$ remains finite. Substituting $\textcolor{red}{\beta}=v/(1-q)$ in the right hand side of Eq. \eqref{stationarity_RS_simple_replica}, and setting $q\to 1^-$ we get
 \begin{equation}
 1=\alpha (1+\sigma^2)\frac{v^2}{(1+v)^2}\Rightarrow v=(\sqrt{\alpha (1+\sigma^2)}- 1)^{-1}\ ,\label{vvsx}
 \end{equation}
 where we have kept only the positive root of the equation, provided that $\alpha>\alpha_c=1/(1+\sigma^2)$.
 Substituting now $\textcolor{red}{\beta}=v/(1-q)$ in the right hand side of Eq. \eqref{Phi_RSb}, setting $q\to 1$, and inserting the result into Eq. \eqref{replica method_intro2v2}, we obtain
 \begin{equation}
 \lim_{\textcolor{ForestGreen}{N}\to\infty}	\frac{\langle  \mathcal E_{min} \rangle_{A,\bm b}}{\textcolor{ForestGreen}{N}} = \frac{\alpha(1+\sigma^2)}{2(1+v)}-\frac{1}{2v}\Big|_{v=(\sqrt{\alpha (1+\sigma^2)}- 1)^{-1}}=\frac{1}{2}\left[\sqrt{\alpha(1+\sigma^2)}-1\right]^2\ ,\label{costfunUnder}
 \end{equation}
 result to be compared with Eq. \eqref{finalEminLagrange} obtained by a completely different method based on Lagrange multipliers and Random Matrix Theory, but only valid for $\alpha>1$.
 \end{enumerate}
 
 To summarise, the quenched replica calculation described in this section and valid for any value $\alpha$ of the ratio between the number $M$ of equations and the number $\textcolor{ForestGreen}{N}$ of the unknowns of a large random linear system $A\bm x=\bm b$ supplemented by the spherical constraint yields the following scenarios depending on the variance $\sigma^2$ of the parameters $b_k$ on the right hand side. For $\alpha<\alpha_c = 1/(1+\sigma^2)<1$, i.e. for ``strongly'' under-complete systems, the average cost function is zero, signalling a system that is typically compatible. For $\alpha>1$, both the Random Matrix and the replica calculations show that over-complete systems are typically incompatible (as the cost function is on average positive, see Eq. \eqref{costfunUnder}) -- which is not surprising, as this would be the case even in absence of the spherical constraint. The most interesting situation happens for $\alpha_c<\alpha<1$, for which the replica calculation shows that such ``weakly'' under-complete systems are also typically incompatible, a new phenomenon entirely due to the extra spherical constraint.

 \subsection{Full Distribution of $\mathcal E_{min}$}\label{sec:fulldist}
 
 In fact, as was observed in \cite{FLD2013}, the replica trick sometimes can be used not only to calculate the mean value of the global minimum but also to characterise fluctuations around it for large $\textcolor{ForestGreen}{N}\gg 1$, employing the Large Deviations approach (for a crash course on large deviations, see Appendix \ref{ref:largedev}). Below we briefly give an account of that idea following the aforementioned paper.

One starts from assuming that the random variable ${\mathcal E}_{min}$ be characterised by a probability density function ${\cal P}_{\textcolor{ForestGreen}{N}}({\mathcal E}_{min})$, which has for large $\textcolor{ForestGreen}{N}\gg 1$ a Large Deviations form
\begin{equation}\label{mean_Large_dev}
{\cal P}_{\textcolor{ForestGreen}{N}}({\mathcal E}_{min})\approx R(x)\mathrm{e}^{-\textcolor{ForestGreen}{N}{\cal L}(x)}, \quad x={\mathcal E}_{min}/\textcolor{ForestGreen}{N}\ .
\end{equation}
Here, we included the leading rate function ${\cal L}(x)$ and also -- for completeness -- a sub-leading pre-factor $R(x)$ that is however not accessible with the method presented here. In \eqref{mean_Large_dev}, the symbol $\approx$ stands for the precise asymptotics\\ $\lim_{\textcolor{ForestGreen}{N}\to\infty}\frac{-1}{\textcolor{ForestGreen}{N}}\log{\cal P}_{\textcolor{ForestGreen}{N}}(\textcolor{ForestGreen}{N} x)={\cal L}(x)$.

On the other hand, consider again the ``replicated'' disorder averaged partition function $\langle Z^{\textcolor{blue}{n}}(\textcolor{red}{\beta})\rangle_V $,
but instead of taking the limits $\textcolor{blue}{n}\to 0$ and $\textcolor{red}{\beta} \to \infty$ separately, perform the double scaling limit $\textcolor{blue}{n}\to 0$, $\textcolor{red}{\beta} \to \infty$ with the product $\textcolor{blue}{n}\textcolor{red}{\beta} =:s$ fixed. In this way, we may write
\begin{equation}\label{LD1}
\lim_{\textcolor{blue}{n}=s/\textcolor{red}{\beta} , \textcolor{red}{\beta} \to \infty}\langle Z^{\textcolor{blue}{n}}(\textcolor{red}{\beta})\rangle_V =\lim_{n=s/\textcolor{red}{\beta} , \textcolor{red}{\beta} \to \infty}\langle \mathrm{e}^{\textcolor{blue}{n}\log Z(\textcolor{red}{\beta})}  \rangle_V=\lim_{\textcolor{red}{\beta} \to \infty}\langle  \mathrm{e}^{\frac{s}{\textcolor{red}{\beta}} \log Z(\textcolor{red}{\beta})} \rangle_V = \langle \mathrm{e}^{-\textcolor{ForestGreen}{N}s x}\rangle_V\ ,
\end{equation}
where in the last step we used the relation \eqref{min_free_energy_intro}. Now we can use the large deviation form \eqref{mean_Large_dev} and rewrite the above as:
\begin{equation}\label{LD2}
\lim_{\textcolor{blue}{n}=s/\textcolor{red}{\beta} , \textcolor{red}{\beta} \to \infty}\langle Z^{\textcolor{blue}{n}}(\textcolor{red}{\beta})\rangle_V  = \langle \mathrm{e}^{-\textcolor{ForestGreen}{N}s x}\rangle_V = \int \mathrm{d}x~ {\cal P}_{\textcolor{ForestGreen}{N}}(x) \mathrm{e}^{-\textcolor{ForestGreen}{N}s x} \approx \int \mathrm{d}x~R(x) \mathrm{e}^{-\textcolor{ForestGreen}{N}\left(s x+{\cal L}(x)\right)}\ ,
\end{equation}
which obviously suggests evaluating the integral by the Laplace method for $\textcolor{ForestGreen}{N}\gg 1$, giving
\begin{equation}\label{LD_Legendre}
\lim_{\textcolor{blue}{n}=s/\textcolor{red}{\beta} , \textcolor{red}{\beta} \to \infty}\langle Z^{\textcolor{blue}{n}}(\textcolor{red}{\beta})\rangle_V \approx g(s)\mathrm{e}^{\textcolor{ForestGreen}{N}\phi(s)}\ ,
\end{equation}
where
\begin{equation}
\phi(s)=-\min_{x} (s x+{\cal L}(x))\ .
\end{equation}
We see that the large deviation rate ${\cal L}(x)$ is related by the so-called Legendre transform
to the function $\phi(s)$ (see \cite{touchette} for an excellent discussion of the Legendre transform in the context of large deviations). Hence, if one could -- by independent means -- find $\phi(s)$, one may recover the rate function ${\cal L}(x)$, governing the large deviation decay of the full probability density ${\cal P}_{\textcolor{ForestGreen}{N}}({\mathcal E}_{min})$ for large $\textcolor{ForestGreen}{N}$
by the inverse Legendre transform:
\begin{equation}\label{LD_Legendre1}
\boxed{{\cal L}(x)=-\left(x s_*+\phi(s_*)\right)}\ ,
\end{equation}
where $s_*:=s_*(x)$ is determined by the following implicit equation
\begin{equation}
 x=-\phi'(s_*)\ .
\end{equation}

 The ability to find the large deviation rate for the minimum of the cost/loss function explicitly hinges on being able to evaluate the above limit  and
to perform the Legendre inversion in closed form. Technically, this is a difficult task, which can be rarely performed successfully. It turns out that our problem is one of these rare and fortunate cases.

Recalling Eq. \eqref{laplaceZQ}
\begin{equation}
\langle Z^{\textcolor{blue}{n}}(\textcolor{red}{\beta})\rangle_V \propto \mathrm{e}^{-\frac{\textcolor{ForestGreen}{N}}{2} \Phi_{\textcolor{blue}{n},\textcolor{red}{\beta}}  (Q_{extr})}\ ,\label{laplaceZQ2}
\end{equation}
and comparing the right hand side with Eq. \eqref{LD_Legendre}, it is clear that
 \begin{equation}\label{LD_replica}
\phi(s)=-\frac{1}{2}\lim_{\textcolor{red}{\beta} \to \infty}\Phi_{\frac{s}{\textcolor{red}{\beta}},\textcolor{red}{\beta}}(Q_{extr})\ ,
\end{equation}
hence we need to evaluate the function $\Phi_{\textcolor{blue}{n},\textcolor{red}{\beta}}(Q_{extr})$
in the limit $\textcolor{blue}{n}\to 0$ and $\textcolor{red}{\beta} \to \infty$ keeping $\textcolor{blue}{n}\textcolor{red}{\beta} =s$ fixed. 

The starting point of this procedure is the expression \eqref{Phi_RSa} for $\Phi_{\textcolor{blue}{n},\textcolor{red}{\beta}}(Q_{extr})$  
 \begin{align}\label{Phi_RSaV2}
\nonumber\Phi_{\textcolor{blue}{n},\textcolor{red}{\beta}} (Q_{extr}) &=\alpha \textcolor{blue}{n}\log{\left(1+\textcolor{red}{\beta} (1-q)\right)}+\alpha\log{\left(1+\frac{\textcolor{red}{\beta} (q+\sigma^2)\textcolor{blue}{n}}{1+\textcolor{red}{\beta} (1-q)}\right)}\\
&-\textcolor{blue}{n}\log{(1-q)}-\log{\left(1+q\frac{\textcolor{blue}{n}}{1-q}\right)}\ ,
\end{align}
as well as the corresponding stationarity condition \eqref{stationarity_RS_simple}
\begin{equation}\label{stationarity_RS_simpleV2}
\frac{q}{(1-q)\left(1+q(\textcolor{blue}{n}-1)\right)}=\alpha \frac{\textcolor{red}{\beta} ^2(q+\sigma^2)}{(1+\textcolor{red}{\beta} -\textcolor{red}{\beta}  q)(1+\textcolor{red}{\beta} -\textcolor{red}{\beta}  q+\textcolor{blue}{n}\textcolor{red}{\beta} (q+\sigma^2))}\ .
\end{equation}

From the previous analysis for $\alpha>\alpha_c$ we expect to have $q\to 1$ as $\textcolor{red}{\beta} \to \infty$ in such a way that $v=\textcolor{red}{\beta} (1-q)$ remains finite\footnote{We do not consider the case $\alpha<\alpha_c$ here, because in that regime we have that $\langle\mathcal{E}_{min}\rangle_{A,\bm b}\to 0$ for large $\textcolor{ForestGreen}{N}$, and $\mathcal{E}_{min}\geq 0$. It follows from Markov's inequality that $\lim_{\textcolor{ForestGreen}{N}\to\infty}\mathrm{Prob}[\mathcal{E}_{min}=0]=1$ for $\alpha<\alpha_c$.}. Correspondingly we substitute $n=s/\textcolor{red}{\beta} , q=1-v/\textcolor{red}{\beta} $ into \eqref{Phi_RSaV2} and set $\textcolor{red}{\beta}  \to \infty$, resulting in the following expression for the functional appearing on the right hand side of \eqref{LD_replica}
\begin{equation}\label{Phi_RS_lim}
\Phi (s,v):= \lim_{\textcolor{red}{\beta} \to \infty}\Phi_{\frac{s}{\textcolor{red}{\beta}},\textcolor{red}{\beta}}(Q_{extr})= \alpha\log{\left(1+\frac{s(1+\sigma^2)}{1+v}\right)}-\log{\left(1+\frac{s}{v}\right)}\ ,
\end{equation}
whereas the stationarity condition  \eqref{stationarity_RS_simpleV2} takes the form
\begin{equation}\label{stationarity_RS_simple2}
\frac{1}{v(v+s)}= \frac{\alpha(1+\sigma^2)}{(1+v)\left(1+v+s(1+\sigma^2)\right)}\ ,
\end{equation}
which is in fact equivalent to the condition $\frac{\partial}{\partial v} \Phi (s,v)=0$.
From this point onwards, one needs to find $v(s)$ solving  \eqref{stationarity_RS_simple2}, which can be equivalently written as
\begin{equation}\label{stationarity_RS_simple3}
\alpha(1+\sigma^2)v(v+s)= {(1+v)\left(1+v+s(1+\sigma^2)\right)}
\end{equation}
and in this way we first get the function $\phi(s)=-\frac{1}{2}\Phi(s,v(s))$ from Eq. \eqref{LD_replica} and \eqref{Phi_RS_lim}. After that, one performs the Legendre transform \label{Legendre transform} over the variable $s$ to obtain the large deviation rate via Eq. \eqref{LD_Legendre1}
\begin{equation}\label{LD_rate_def1}
{\cal L}(x)=-x s_*+\frac{1}{2}\Phi\left(s_*,v(s_*)\right)=\frac{1}{2}\left[\Phi\left(s_*,v(s_*)\right)-2 x s_*\right],
\end{equation}
where $s_*$ as a function of $x$ is found by solving the equation
\begin{equation}\label{LD_rate_def2}
x=\left. \frac{1}{2}\frac{\mathrm{d}\Phi}{\mathrm{d}s}\right\vert_{s_*}  = \left. \frac{1}{2}\frac{\partial\Phi(s,v)}{\partial s} \right\vert_{s_*},
\end{equation}
where the total derivative of $\Phi$ with respect to $s$ coincides with the partial derivative because of the above-mentioned stationarity $\frac{\partial}{\partial v} \Phi (s,v)=0$ with respect to any indirect dependence on $s$ through $v(s)$.

Differentiating \eqref{Phi_RS_lim} we find, using \eqref{stationarity_RS_simple2}, that
\begin{equation}\label{LD_rate_def3}
x=\frac{1}{2}\left(\frac{\alpha(1+\sigma^2)}{1+v+s_*(1+\sigma^2)}-\frac{1}{v+s_*}\right)=
\frac{1}{2}\left(\frac{1+v}{v(v+s_*)}-\frac{1}{v+s_*}\right)=\frac{1}{2v(v+s_*)}\ ,
\end{equation}
or equivalently
\begin{equation}\label{LD_rate_def4}
v(v+s_*)=\frac{1}{2 x}\ .
\end{equation}
Using the above, one can rewrite
\eqref{stationarity_RS_simple3} as
\begin{equation}\label{stationarity_RS_simple4}
\alpha \left(1+\sigma^2 \right) \frac{1}{2 x}= {(1+v) \left(1+v+s_*(1+\sigma^2)\right) }\ .
\end{equation}
Further expressing $s_*$ from \eqref{stationarity_RS_simple3} as
\begin{equation}\label{stationarity_RS_simple5}
s_*(1+\sigma^2)=\frac{v^2\left(\alpha(1+\sigma^2)-1\right)-2v-1}{1-v(\alpha-1)}
\end{equation}
and substituting back into \eqref{stationarity_RS_simple4} we get the closed-form equation for $v$ as a function of
$x$
\begin{align}
\nonumber \alpha(1+\sigma^2)\frac{1}{2x} &= (1+v)\left\{1+v+\frac{v^2\left(\alpha(1+\sigma^2)-1\right)-2v-1}{1-v(\alpha-1)}\right\}\\
&=
\frac{\alpha(1+v)v(v\sigma^2-1)}{1-v(\alpha-1)}\ , \label{LD_rate_def5}
\end{align}
which can be equivalently rewritten as a cubic equation 
\begin{equation}\label{LD_rate_def6}
\boxed{\sigma^2v^3+v^2(\sigma^2-1)-v\left(1-a(\alpha-1)\right)-a=0\ , \quad a:=\frac{1+\sigma^2}{2x}}\ .
\end{equation}
Solving this equation, one gets the value of $v$ corresponding to a given $x$.
It turns out that the knowledge of $v(x)$ is enough to get the large deviation rate $\mathcal{L}(x)$, which can be expressed solely via such $v$. The simplest way to proceed is by multiplying Eq.  \eqref{stationarity_RS_simple2}  
by $(1+v)^2$, taking logarithms of both sides, and further multiplying by $\alpha$ to get
\begin{equation}\label{station_ident}
\alpha\log{\left(\frac{1+v+s(1+\sigma^2)}{1+v}\right)}=\alpha  \log{\left(\frac{v(v+s)}{(1+v)^2}\alpha(1+\sigma^2)\right)}\ ,
\end{equation}
which allows to rewrite the expression \eqref{Phi_RS_lim} first as
\begin{equation}\label{Phi_RS_lim2}
\Phi (s,v)= \alpha  \log{\left(\frac{v(v+s)}{(1+v)^2}\alpha(1+\sigma^2)\right)}-\log{\left(v(v+s)\right)}+2\log{v}\ ,
\end{equation}
and then using \eqref{LD_rate_def4} as
\begin{equation}\label{Phi_RS_lim2}
\Phi (s_*(x),v)=-(\alpha-1)\log{(2x)}+\alpha\log{\left(\alpha(1+\sigma^2)\right)}+2\log v-2\alpha\log{(1+v)}\ .
\end{equation}
This should be combined with the relation
\begin{equation}
xs_*=\frac{1}{2v}-xv\label{eqxsv}
\end{equation}
following from \eqref{LD_rate_def4}. In this way, one completely solves the problem of explicitly inverting the Legendre transform and providing the required  Large Deviation Rate function \eqref{LD_rate_def1} for the minimal cost as a function of $x$.

\begin{figure}[htb]
\begin{center}
\includegraphics[scale=0.25,clip=true]{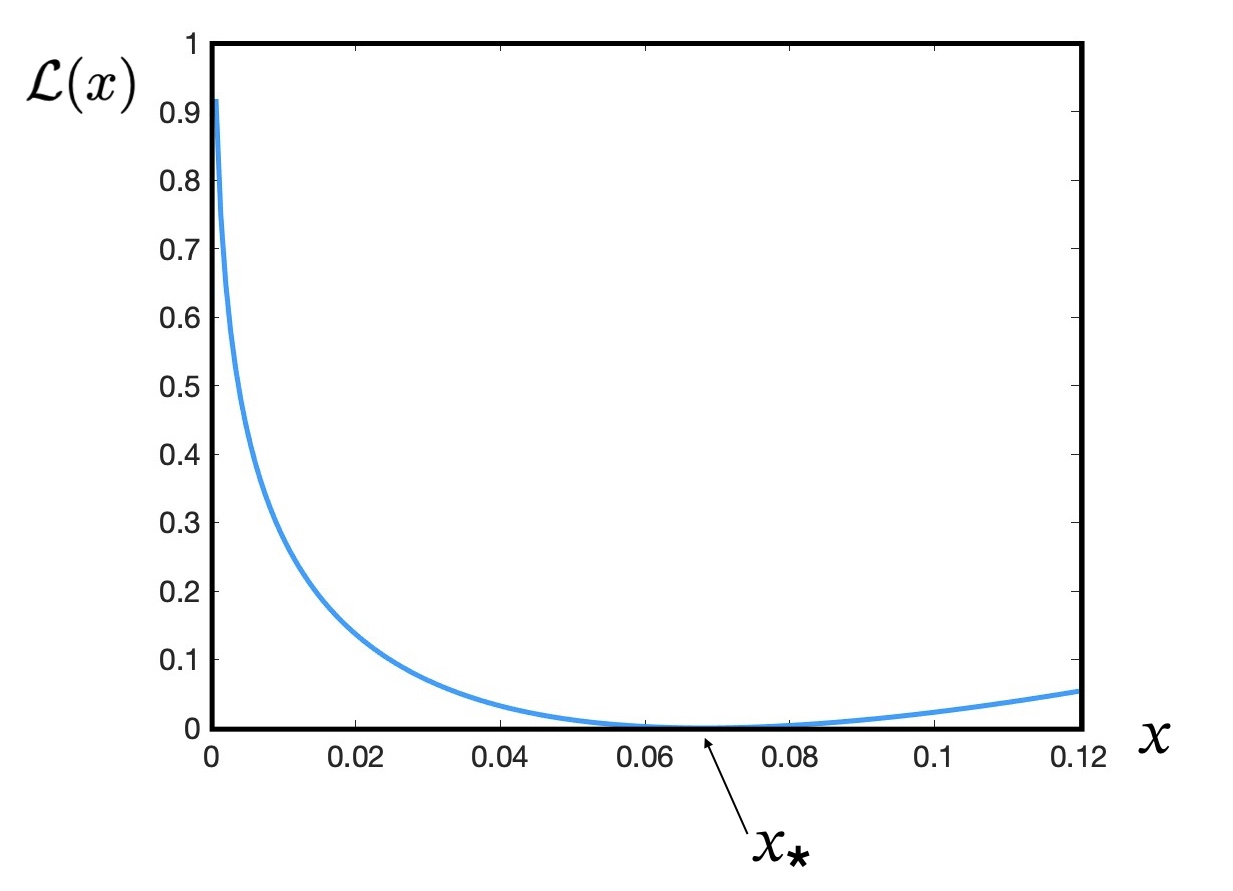}
\caption{Rate function $\mathcal{L}(x)$ for $\alpha=1.5$ and $\sigma=0.5$, obtained numerically starting from the cubic equation Eq. \eqref{LD_rate_def6} and using the procedure detailed above.} \label{figRate}
\end{center}
\end{figure}

To summarise the operational steps, one first fixes the requested variance of the noise terms $\sigma^2$ and the ratio $\alpha>\alpha_c=1/(1+\sigma^2)$ between the number of equations and the number of unknowns. Then, one constructs a mesh $\mathcal{M}=\{x_1,\ldots,x_K\}$ of $x>0$ values, which should include the special value 
\begin{equation}
x_*:=\frac{1}{2}\left[\sqrt{\alpha(1+\sigma^2)}-1\right]^2\label{xstarRep}
\end{equation}
for which we know that the corresponding value of $v$ must be $v_*=1/\sqrt{2x_*}$ (see Eq. \eqref{vvsx} and Eq. \eqref{costfunUnder}). This piece of information is useful to select the corresponding root of the cubic equation in Eq. \eqref{LD_rate_def6}. Following by continuity the solution of the cubic equation along the $x$-mesh $\mathcal{M}$, one construct a set of corresponding $v$-values $\{v_1,\ldots, v_K\}$. Using Eq. \eqref{eqxsv}, one can then construct the corresponding set of $s_*$ values, $\{(s_*)_1,\ldots, (s_*)_K\}$. Finally, invoking Eq. \eqref{LD_rate_def1} and \eqref{Phi_RS_lim2}, we get the values of the rate function corresponding to the $x$-mesh points as
\begin{align}
\nonumber \mathcal{L}(x_j) &= -x_j (s_*)_j+\frac{1}{2}\left[-(\alpha-1)\log{(2x_j)}+\alpha\log{\left(\alpha(1+\sigma^2)\right)}+2\log v_j\right.\\
&\left.-2\alpha\log{(1+v_j)}\right]
\end{align}
for $j=1,\ldots,K$. The rate function $\mathcal{L}(x)$ is plotted in Fig. \ref{figRate}. However, determining the precise domain of validity of $\mathcal{L}(x)$ away from its minimum $x_*$ is a difficult problem, which is still largely unsolved, and for which a mathematically rigorous treatment of this problem is very much called for. What can be computed from the expression for $\mathcal{L}(x)$, though, is the quadratic behaviour around its minimum, which provides the variance of $\mathcal{E}_{min}$ in the regime of typical fluctuations.

Combining Eqs. \eqref{Phi_RS_lim2},  \eqref{eqxsv} and \eqref{LD_rate_def1}, we can write
\begin{align}
\nonumber\mathcal{L}(x,v(x)) &= \frac{1}{2} \left(\alpha  \log \left(\alpha  \left(\sigma ^2+1\right)\right)-2 \alpha  \log (v(x)+1)+2 \log (v(x))\right.\\
&\left.+(1-\alpha ) \log (2 x)\right)+x v(x)-\frac{1}{2 v(x)}\ .
\end{align}
Taylor-expanding the rate function up to second order around $x=x_*$ (see Eq. \eqref{xstarRep}), we get
\begin{align}
\nonumber\mathcal{L}(x,v(x)) &=\mathcal{L}(x_*,v(x_*))  +(x-{x_*})f_1(x_*)+(x-{x_*})^2 f_2(x_*) +\mathcal{O}\left((x-{x_*})^3\right)\ ,
\end{align}
where
\begin{align}
f_1(x_*)&= \left(\frac{1}{2} \left(-\frac{2 \alpha  v'({x_*})}{v({x_*})+1}+\frac{2 v'({x_*})}{v({x_*})}+\frac{1-\alpha }{{x_*}}\right)+{x_*} v'({x_*})+\frac{v'({x_*})}{2 v({x_*})^2}+v({x_*})\right)\\
\nonumber f_2(x_*) &=\left[\frac{1}{2} \left(\alpha  \left(\frac{v'({x_*})^2}{(v({x_*})+1)^2}-\frac{v''({x_*})}{v({x_*})+1}\right)+\frac{v({x_*}) v''({x_*})-v'({x_*})^2}{v({x_*})^2}+\frac{\alpha -1}{2 {x_*}^2}\right)\right.\\
&\left.+\frac{v({x_*}) v''({x_*})-2 v'({x_*})^2}{4 v({x_*})^3}+v'({x_*})+\frac{1}{2} {x_*} v''({x_*})\right]\ .
\end{align}
The first and second derivative of $v(x)$, evaluated at $x_*$, can be computed differentiating Eq. \eqref{LD_rate_def6} once and twice with respect to $x$, respectively. After some tedious algebra, we find that $\mathcal{L}(x_*,v(x_*)) =f_1(x_*)=0$ as expected (as the rate function should have
a minimum (a zero) at the minimal expected value of $\mathcal{E}_{min}/N=x_*$), while the second-order term reads
\begin{equation}
f_2(x_*)=\frac{1}{\sigma ^2 \left(\sqrt{\alpha  \left(1+\sigma ^2\right)}-1\right)^2}=\frac{1}{2x_*\sigma^2}\ .
\end{equation}
Inserting this result into Eq. \eqref{mean_Large_dev}, we find that
\begin{align}
\nonumber {\cal P}_N({\mathcal E}_{min}) &\approx \exp\left[-N \frac{1}{2x_*\sigma^2}\left(\frac{\mathcal{E}_{min}}{N}-x_*\right)^2\right]\\
&= \exp\left[-\frac{1}{2(x_*\sigma^2 N)}\left(\mathcal{E}_{min}-Nx_*\right)^2\right]\ ,
\end{align}
corresponding to Gaussian typical fluctuations of $\mathcal{E}_{min}$ for large $N$ and $\alpha>\alpha_c$ around its average value $N x_*$, and with variance
\begin{equation}
\mathrm{Var}(\mathcal{E}_{min})=x_*\sigma^2 N\ .
\end{equation}

 \section{Conclusions and Perspectives}
 
 These lecture notes constitute an attempt to illustrate and advertise a series of works by Yan V. Fyodorov and Rashel Tublin on a rich and intriguing problem, namely the statistical study of solutions of linear systems in presence of an additional quadratic constraint \cite{Yan1,Yan2,tublin}. This problem is interesting especially for its pedagogical and instructive nature: it can be tackled with two largely complementary methods, and lends itself to a number of interesting generalisations and follow-ups, most of which have not been tackled yet.
 
 The first method we discussed in these notes starts from the definition of the cost function $H(\bm x)$ in Eq. \eqref{defHx}, which is zero when the linear system is compatible, and larger than zero when it is incompatible. Using one Lagrange multiplier to enforce the spherical constraint on the solution vector, $||\bm x||^2=\textcolor{ForestGreen}{N}$, one may determine the location of the critical points of $H(\bm x)$ (see Eq. \eqref{defX}), as well as the equation for the Lagrange multiplier (see Eqs. \eqref{sphconstraint} and \eqref{spherical2}). In a random/disordered setting -- where the matrix $A$ and the vector of coefficients $\bm b$ are Gaussian variables -- the average of the minimal Lagrange multiplier can be computed exactly for large $\textcolor{ForestGreen}{N}$ and $M$, with $M/\textcolor{ForestGreen}{N}=\alpha>1$ fixed (see Eq. \eqref{lambdastar}). In turn, this result can be used -- under a self-averaging assumption -- to compute the average $\langle\mathcal{E}_{min}\rangle_{A,\bm b}$ of the minimal cost, which signals whether the corresponding linear system is expected to be typically compatible or incompatible. While in the absence of the spherical constraint, over-complete systems ($\alpha>1$) are typically incompatible, and under-complete systems ($\alpha<1$) are typically compatible (see Appendix \ref{appnospherical} for a detailed calculation), the spherical constraint induces a further intermediate regime for $0<\alpha_c<\alpha<1$ where ``weakly'' under-complete systems may still be typically incompatible, and only for ``strongly'' under-complete systems the presence of the spherical constraint does not typically hinder the possibility to find a solution.
 
 These findings are independently confirmed using an entirely different line of reasoning, inspired by an intriguing connection between the search for the average of the minimum of a random variable, and the free energy of an associated thermodynamical system at zero temperature. Computing the quenched average of the free energy using the celebrated \emph{replica trick} from the physics of disordered systems yields an independent confirmation of a transition between compatible and incompatible linear systems, taking place at a value $\alpha_c<1$, which can be precisely characterised (see Eq. \eqref{alphacreplicas}). As a further (and rare) luxury afforded to us by this problem, the replica calculation in this case allows us access not only to the average of the minimal cost $\langle\mathcal{E}_{min}\rangle_{A,\bm b}$, but also to its full distribution $\mathcal{P}_{\textcolor{ForestGreen}{N}}(\mathcal{E}_{min})$ (in a large deviation sense) -- see discussion in Section \ref{sec:fulldist}. 
 
 Given the heuristic nature of the replica calculation, a mathematically rigorous justification of the steps contained in Section \ref{sec:replicas} is still lacking. It is worth mentioning, though, that many questions related to a simpler optimisation problem on the sphere studied originally in \cite{FLD2013} by a combination of rigorous Kac-Rice and heuristic replica approaches were subsequently successfully put on firm mathematical grounds in the series of papers \cite{DemZeit,Kivimae,Sosoe20,Baiketal21,Belius21}, therefore it is likely that some of those techniques can be also useful in the context of present model as well.

 In terms of future perspectives, some interesting generalisations and follow-ups come to mind:
 \begin{enumerate}
 \item The function $f(u)=\sigma^2+u$ in Eq. \eqref{deffu} is specific to the linear nature of the equations of the system. Other choices are possible, for instance a quadratic $f(u)$ would correspond to a system of quadratic equations with a spherical constraint. This extended class of problems has been addressed only very recently \cite{urbanirig,urb1,urb2}, with a number of exciting results and conjectures, and intriguing applications for example to the so-called ``phase retrieval" problem, see \cite{phaseretrieval1,urb3}.

\item Lifting the restriction that both $A$ and $\bm b$ be Gaussian distributed would be very welcome. For instance, what is the effect on the compatibility threshold of having ``fat-tailed'' coefficients of the linear system? Relinquishing Gaussianity is always tricky, though, especially in the context of replica calculations, where the evaluation of the average over the disorder (see \eqref{averageV} and \eqref{averageV2}) may then lead to a multivariate integral much less amenable to a Laplace (saddle-point) evaluation. Overcoming this technical hurdle is likely to open up an interesting new field of investigations.
\item The quest for a rigorous -- or at least, less non-rigorous -- way to perform the replica calculation in Section \ref{sec:replicas} may actually require letting Random Matrix Theory back in the game. The idea would be to attempt an exact evaluation of the integral \eqref{<Z^n> pre} for any $\textcolor{ForestGreen}{N}$ and $\textcolor{blue}{n}$, without any approximation. The key idea -- pioneered by Kanzieper \cite{kanzieper} and later implemented in the context of spherical spin glasses by Akhanjee and Rudnick \cite{rudnick} -- would be to diagonalise the symmetric matrix $Q$ as $Q=OMO^{-1}$, with $M$ the diagonal matrix containing the $\textcolor{blue}{n}$ eigenvalues $\{\mu_1,\ldots,\mu_{\textcolor{blue}{n}}\}$ and $O$ the orthogonal matrix of eigenvectors, and use the fact that the integrand of \eqref{<Z^n> pre} is ``almost'' rotationally invariant, as it depends only weakly on the eigenvectors of $Q$. Reducing the $Q$ integral to a $\textcolor{blue}{n}$-fold integral over the eigenvalues of $Q$ could prelude an exact evaluation of the replicated partition function for any $\textcolor{ForestGreen}{N}$ and $\textcolor{blue}{n}$ in terms of Painlev\'e transcendents. This research programme could shed light on the interplay between replica symmetry (or breaking thereof) and spectral properties of the replicated partition function without any large $\textcolor{ForestGreen}{N}$ approximation.
 \end{enumerate}

\newpage
\appendix

\section{Appendix: Minimal cost function without spherical constraint}\label{appnospherical}

In this Appendix, we wish to show how the spherical constraint is indeed the sole responsible for the new phenomenon of ``weakly under-complete'' systems, by computing the average minimal loss function in the absence of such constraint. The calculation can be performed exactly without replicas.

The partition function Eq. \eqref{method Z_intro} for a single instance of the ``disorder'' (randomness) encoded in $A$ and $\bm b$ reads in this case
\begin{equation}\label{method Z_Appendix1_nospheric}
		Z(\textcolor{red}{\beta}) = \int_{\mathbb{R}^{\textcolor{ForestGreen}{N}}}\mathrm{d} \bm x~\mathrm{e}^{-\frac 12 \textcolor{red}{\beta}   \sum_{k=1}^M [\sum_{j=1}^{\textcolor{ForestGreen}{N}} A_{kj}x_j-b_k]^2}\ ,
	\end{equation}
where we defer considerations about the convergence of this integral to a later stage, and we did not include any spherical constraint here.

Expanding the exponent, we have
\begin{align}
\nonumber &-\frac 12 \textcolor{red}{\beta}   \sum_{k=1}^M \left[\sum_{j=1}^{\textcolor{ForestGreen}{N}} A_{kj}x_j-b_k\right]^2=-\frac 12 \textcolor{red}{\beta}   \sum_{k=1}^M \left[\sum_{j=1}^{\textcolor{ForestGreen}{N}} A_{kj}x_j-b_k\right]
\left[\sum_{\ell=1}^{\textcolor{ForestGreen}{N}} A_{k\ell}x_\ell-b_k\right]\\
\nonumber &=-\frac 12 \textcolor{red}{\beta} \left[ \sum_{j,\ell}x_i \left(\sum_k A_{kj}A_{k\ell}\right)x_\ell-\sum_j \left(2\sum_k A_{kj}b_k\right)x_j+\sum_k b_k^2\right]\\
&=-\frac{1}{2}\bm x^T W_{\textcolor{red}{\beta}}\bm x+\bm h^T_{\textcolor{red}{\beta}}\bm x-\frac{1}{2}\textcolor{red}{\beta}\sum_k b_k^2\ ,
\end{align}
where
\begin{align}
W_{\textcolor{red}{\beta}} &=\textcolor{red}{\beta}A^T A\\
\bm h_{\textcolor{red}{\beta}} &=\textcolor{red}{\beta}A^T \bm b\ .
\end{align}
Hence, for the integral \eqref{method Z_Appendix1_nospheric} to converge, we need the matrix $W_{\textcolor{red}{\beta}}$ to be positive definite, which requires the $\textcolor{ForestGreen}{N}\times \textcolor{ForestGreen}{N}$ matrix $A^T A$ to be of Wishart (and not of anti-Wishart) type. This happens if $M>\textcolor{ForestGreen}{N}$ ($\alpha>1$). The case $M<\textcolor{ForestGreen}{N}$ -- for which the matrix $A^T A$ has zero eigenvalues -- must be treated separately (see below).

Therefore, for $\alpha>1$ we have
\begin{align}
\nonumber Z(\textcolor{red}{\beta}) &=\exp\left[-\frac{1}{2}\textcolor{red}{\beta}\sum_k b_k^2\right] \int_{\mathbb{R}^{\textcolor{ForestGreen}{N}}}\mathrm{d} \bm x~\mathrm{e}^{-\frac{1}{2}\bm x^T W_{\textcolor{red}{\beta}}\bm x+\bm h^T_{\textcolor{red}{\beta}}\bm x}\\
&=\sqrt{\frac{(2\pi)^{\textcolor{ForestGreen}{N}}}{\det W_{\textcolor{red}{\beta}}}}\exp\left[-\frac{1}{2}\textcolor{red}{\beta}\sum_k b_k^2\right] \exp\left[\frac{1}{2}\bm h^T_{\textcolor{red}{\beta}} W_{\textcolor{red}{\beta}}^{-1}\bm h_{\textcolor{red}{\beta}}\right]\ .
\end{align}
Recalling formula \eqref{min_free_energy_intro}
	\begin{equation}\label{min_free_energy_appendix}
		 \mathcal E_{min}  = -\lim_{\textcolor{red}{\beta} \to\infty}\frac{1}{\textcolor{red}{\beta} }  \log Z(\textcolor{red}{\beta}) \ ,
	\end{equation}
we have explicitly
	\begin{equation}\label{min_free_energy_appendix_expl}
		 \mathcal E_{min}  = \frac{1}{2}\sum_{k=1}^M b_k^2-\frac{1}{2}\bm b^T A(A^T A)^{-1}A^T \bm b\ .
	\end{equation}
	
We can now compute the average of $ \mathcal E_{min} $ over $\bm b$ first, using the identities \eqref{id1} and \eqref{id2}
	\begin{align}
		\langle \mathcal E_{min}\rangle_{\bm b} & = \frac{1}{2}M\sigma^2-\frac{1}{2}\sigma^2\mathrm{Tr}\left(A(A^T A)^{-1}A^T\right) \\
		 &=\frac{1}{2}M\sigma^2-\frac{1}{2}\sigma^2\mathrm{Tr}\mathbb{I}_{\textcolor{ForestGreen}{N}}=\frac{1}{2}\sigma^2 (M-\textcolor{ForestGreen}{N})\ , \label{min_free_energy_appendix_expl_avb}
	\end{align}
	where we used the cyclic property of the trace. Interestingly, the result holds irrespective of what the $A$ matrix is -- a clear indication that the nature of the coefficient matrix is irrelevant in determining whether a linear system with more equations than unknowns is compatible (it typically isn't). Therefore, in the large $\textcolor{ForestGreen}{N}$ limit, we can write
	\begin{equation}
	\lim_{\textcolor{ForestGreen}{N}\to\infty} \frac{\langle \mathcal E_{min}\rangle_{A,\bm b} }{\textcolor{ForestGreen}{N}}=\frac{1}{2}\sigma^2 (\alpha-1)\ ,
	\end{equation}
	which is indeed positive for $\alpha>1$, signalling that an over-complete system of equations (even in the absence of the spherical constraint) is typically incompatible, as it should.

For $\alpha<1$ (more unknowns than equations), the integral defining the partition function \eqref{method Z_Appendix1_nospheric} does not converge, but it can be regularised as follows
\begin{equation}\label{method Z_Appendix1_nospheric}
		Z_\lambda(\textcolor{red}{\beta}) = \int_{\mathbb{R}^{\textcolor{ForestGreen}{N}}}\mathrm{d} \bm x~\mathrm{e}^{-\frac 12 \textcolor{red}{\beta}   \sum_{k=1}^M [\sum_{j=1}^{\textcolor{ForestGreen}{N}} A_{kj}x_j-b_k]^2-\lambda \sum_{j=1}^{\textcolor{ForestGreen}{N}} x_j^2}\ ,
	\end{equation}
which has the effect of penalising solutions $\bm x$ with large norms for $\lambda>0$ -- it is essentially a ``soft'' version of the ``hard'' spherical constraint used in the main text. This method is called \emph{$L_2$-regularisation} \cite{boyd}.

Therefore, for $\alpha<1$ we can write
\begin{align}
\nonumber Z_\lambda(\textcolor{red}{\beta}) &=\exp\left[-\frac{1}{2}\textcolor{red}{\beta}\sum_k b_k^2\right] \int_{\mathbb{R}^{\textcolor{ForestGreen}{N}}}\mathrm{d} \bm x~\mathrm{e}^{-\frac{1}{2}\bm x^T W_{\textcolor{red}{\beta},\lambda}\bm x+\bm h^T_{\textcolor{red}{\beta}}\bm x}\\
&=\sqrt{\frac{(2\pi)^{\textcolor{ForestGreen}{N}}}{\det W_{\textcolor{red}{\beta},\lambda}}}\exp\left[-\frac{1}{2}\textcolor{red}{\beta}\sum_k b_k^2\right] \exp\left[\frac{1}{2}\bm h^T_{\textcolor{red}{\beta}} W_{\textcolor{red}{\beta},\lambda}^{-1}\bm h_{\textcolor{red}{\beta}}\right]\ ,
\end{align}
where 
\begin{equation}
W_{\textcolor{red}{\beta},\lambda} =\textcolor{red}{\beta}A^T A+2\lambda\mathbb{I}_{\textcolor{ForestGreen}{N}}\ ,
\end{equation}
which is positive definite, being the sum of a positive semi-definite matrix (the `anti-Wishart' matrix $\textcolor{red}{\beta}A^T A$ \cite{antiwishart1,antiwishart2}) and a positive definite matrix (a positive multiple of the identity).

Recalling formula \eqref{min_free_energy_intro} again -- suitably deformed to remove the regulariser 
	\begin{equation}\label{min_free_energy_appendix_again}
		 \mathcal E_{min}  = -\lim_{\lambda\to 0}\lim_{\textcolor{red}{\beta} \to\infty}\frac{1}{\textcolor{red}{\beta} }  \log Z_\lambda(\textcolor{red}{\beta}) \ ,
	\end{equation}
we have explicitly
	\begin{equation}\label{min_free_energy_appendix_expl_again}
		 \mathcal E_{min}  = \frac{1}{2}\sum_{k=1}^M b_k^2-\frac{1}{2}\lim_{\lambda\to 0}\lim_{\textcolor{red}{\beta}\to\infty}\textcolor{red}{\beta}\bm b^T A(\textcolor{red}{\beta}A^T A+2\lambda\mathbb{I}_{\textcolor{ForestGreen}{N}})^{-1}A^T \bm b\ .
	\end{equation}
	
We can now compute the average of $ \mathcal E_{min} $ over $\bm b$ first, using the identities \eqref{id1} and \eqref{id2}
	\begin{align}
		\langle \mathcal E_{min}\rangle_{\bm b} & = \frac{1}{2}M\sigma^2-\frac{1}{2}\sigma^2\lim_{\lambda\to 0}\lim_{\textcolor{red}{\beta}\to\infty}\textcolor{red}{\beta}\mathrm{Tr}\left((\textcolor{red}{\beta}A^T A+2\lambda\mathbb{I}_{\textcolor{ForestGreen}{N}})^{-1}A^T A\right)\ ,\label{min_free_energy_appendix_expl_avb_again}
	\end{align}
	where we used the cyclic property of the trace again. The average over the matrix $A$ can be performed recalling that the $\textcolor{ForestGreen}{N}\times \textcolor{ForestGreen}{N}$ anti-Wishart matrix $W^{(a)}=A^T A$ for $M<\textcolor{ForestGreen}{N}$ has the following density of eigenvalues for large $\textcolor{ForestGreen}{N},M$ with fixed ratio $\alpha=M/\textcolor{ForestGreen}{N}<1$
	\begin{equation}
	\rho(s) = (1-\alpha)\delta(s)+\alpha \rho_{MP}(s)\ ,\label{rhoAW}
	\end{equation}
 which signals the fact that $W^{(a)}=A^T A$ has exactly $\textcolor{ForestGreen}{N}-M$ eigenvalues equal to zero (see Eq. \eqref{rhoAW1}). Therefore
\begin{equation}
\Big\langle\frac{1}{\textcolor{ForestGreen}{N}} \mathrm{Tr}\left((\textcolor{red}{\beta}W^{(a)}+2\lambda\mathbb{I}_{\textcolor{ForestGreen}{N}})^{-1}W^{(a)}\right)\Big\rangle_{W^{(a)}}\to\alpha\int_{s_-}^{s_+}\mathrm{d}s\frac{s}{\textcolor{red}{\beta}s+2\lambda}\rho_{MP}(s)\ ,
\end{equation}	
where we have used the property \eqref{traceproperty} to compute the average trace of a matrix as an integral over its eigenvalue density, and observed that the Dirac delta term in Eq. \eqref{rhoAW} does not contribute to the integral.

Multiplying the right hand side up and down by $\textcolor{red}{\beta}$, and adding and subtracting $2\lambda$ in the integrand, we get
\begin{align}
\nonumber\Big\langle\frac{1}{\textcolor{ForestGreen}{N}} \mathrm{Tr}\left((\textcolor{red}{\beta}W^{(a)}+2\lambda\mathbb{I}_{\textcolor{ForestGreen}{N}})^{-1}W^{(a)}\right)\Big\rangle_{W^{(a)}} &\to\frac{\alpha}{\textcolor{red}{\beta}}\int_{s_-}^{s_+}\rho_{MP}(s)\\
& -\frac{2\lambda\alpha}{\textcolor{red}{\beta}}\int_{s_-}^{s_+}\mathrm{d}s\frac{1}{\textcolor{red}{\beta}s+2\lambda}\rho_{MP}(s)\ .
\end{align}		
Clearly, the first term on the right hand side is equal to $\alpha/\textcolor{red}{\beta}$ since the Mar\v cenko-Pastur density is normalised to one, while the second term is of order $1/\textcolor{red}{\beta}^2$ for large $\textcolor{red}{\beta}$ at fixed $\lambda$, and therefore will not contribute once multiplied by an extra $\textcolor{red}{\beta}$ in Eq. \eqref{min_free_energy_appendix_expl_avb_again}.
	
Therefore, in the large $\textcolor{ForestGreen}{N}$ limit, we can write from Eq. \eqref{min_free_energy_appendix_expl_avb_again}
	\begin{equation}
	\lim_{\textcolor{ForestGreen}{N}\to\infty} \frac{\langle \mathcal E_{min}\rangle_{A,\bm b} }{\textcolor{ForestGreen}{N}}=\frac{1}{2}\sigma^2 \alpha - \frac{1}{2}\sigma^2 \alpha=0\ ,
	\end{equation}
signalling that an under-complete system of equations is typically compatible, as it should.

\section{Appendix: Generalities on Wishart ensemble}\label{app:Wishart}

In this Appendix, we collect some salient facts about the Wishart ensemble of random matrices.

\begin{enumerate}
\item If $A_{ij}$ are i.i.d. real Gaussian variables with mean zero and variance $1/N$, $P_A(A)=\prod_{i=1}^M \prod_{j=1}^N p(A_{ij}=x)$, where $p(A_{ij}=x)=\exp(-Nx^2/2)/\sqrt{2\pi/N}$, with $M>N$, then the $N\times N$ symmetric matrix
$W = A^T A$ belongs to the Wishart ensemble. 
\item The entries of $W$ are no longer independent, and their joint probability density (jpd) $P_W(W):=P(W_{11},\ldots,W_{NN})$ is
\begin{equation}
P_W(W)=C_{N,M}(\det W)^{\frac{M-N-1}{2}}\mathrm{e}^{-\frac{N}{2}\mathrm{Tr} W}\ ,\label{entries}
\end{equation}
where the normalisation factor $C_{N,M}$ is known in closed form.
\item The $N$ eigenvalues $\{s_1,\ldots,s_N\}$ of $W$ are real and non-negative. Their jpd $P(\bm s)$ follows from \eqref{entries} as
\begin{equation}
P(\bm s)=\hat C_{N,M}\mathrm{e}^{-\frac{N}{2}\sum_{k=1}^N s_k}\prod_{k=1}^N s_k^{\frac{M-N-1}{2}}|\Delta(\bm s)|\ ,\label{eigenvalues}
\end{equation}
where $\Delta(\bm s)=\prod_{j<k}(s_j-s_k)$ -- the Vandermonde determinant -- is the Jacobian of the change of variables $W\to OS O^{-1}$, where $O$ is the orthogonal matrix formed by the eigenvectors of $W$, and $S$ is the diagonal matrix of eigenvalues.
\item The density of eigenvalues $\rho_N(s)=\Big\langle\frac{1}{N} \sum_{i=1}^N \delta(s-s_i)\Big\rangle$ (normalised to $1$) converges for large $N$ and $M$, with $M/N=\alpha>1$, to the Mar\v cenko-Pastur law
\begin{equation}
\lim_{N\to\infty} \rho_N(s) = \rho_{MP}(s)\ ,
\end{equation}
given by Eq. \eqref{MPeq}
\begin{equation}
\rho_{MP}(s)=\frac{2}{\pi s}\frac{1}{(\sqrt{s_+}-\sqrt{s_-})^2}\sqrt{(s_+-s)(s-s_-)}
\end{equation}
having compact support between the edge points $s_\pm=(\sqrt{\alpha}\pm 1)^2$ on the positive real axis (see Fig. \ref{figMP}). The density of eigenvalues $\rho_N(s)$ for finite $N$ is known in closed form \cite{livanRMT}.
\item The \emph{anti-Wishart} $M\times M$ matrix $W^{(a)}=AA^T$ \cite{antiwishart1,antiwishart2} shares the same set of non-negative eigenvalues $\{s_i\}_{i=1,\ldots,N}$ with $A$, plus $M-N$ exactly zero eigenvalues. \textit{Proof:} if $A^T A\bm x_i=s_i\bm x_i$, multiply to the left by $A$. The spectral density of the anti-Wishart ensemble is therefore given by the Mar\v cenko-Pastur law, supplemented by a delta function at zero eigenvalue, i.e.
	\begin{equation}
	\rho(s) = (1-\alpha)\delta(s)+\alpha \rho_{MP}(s)\ .\label{rhoAW1}
	\end{equation}
\end{enumerate}

\section{Appendix: Sketch of Browne's theorem}\label{appBrowne}

We sketch here the main gist of the proof of Browne's theorem \cite{Browne1967}, essentially identical to Gander's \cite{Gander1981}.

Consider two distinct solutions $\bm x_1$ and $\bm x_2$ of the stationarity condition \eqref{Lagr1}
\begin{align}
 (A^T A-\lambda_1\mathbb{I}_N)\bm x_1 &=A^T\bm b\ ,\label{Lagr1app1}\\
  (A^T A-\lambda_2\mathbb{I}_N)\bm x_2 &=A^T\bm b\ ,\label{Lagr1app2}
 \end{align}
 corresponding to two distinct Lagrange multipliers $\lambda_1$ and $\lambda_2$. We aim to compute the following difference
 \begin{equation}
 \Delta(\bm x_1,\bm x_2;\lambda_1,\lambda_2)=H(\bm x_2)-H(\bm x_1)=||A\bm x_2-\bm b||^2-||A\bm x_1-\bm b||^2\ .\label{Deltadef}
 \end{equation}
First, multiply Eq. \eqref{Lagr1app1} by $\bm x_1^T$ to the left, and Eq. \eqref{Lagr1app2} by $\bm x_2^T$ to get
\begin{align}
||A\bm x_1||^2-\bm x_1^T A^T \bm b &=\lambda_1 ||\bm x_1||^2\\
||A\bm x_2||^2-\bm x_2^T A^T \bm b &=\lambda_2 ||\bm x_2||^2\ .
\end{align}
Transposing the above equations
\begin{align}
||A\bm x_1||^2-\bm b^T A \bm x_1 &=\lambda_1 ||\bm x_1||^2\\
||A\bm x_2||^2-\bm b^T A \bm x_2 &=\lambda_2 ||\bm x_2||^2\ ,
\end{align}
and subtracting the first from the second, we get
\begin{equation}
||A\bm x_2||^2-||A\bm x_1||^2 -\bm b^T A (\bm x_2-\bm x_1)=\lambda_2 ||\bm x_2||^2-\lambda_1 ||\bm x_1||^2\ .\label{sum1}
\end{equation}

Now, multiply Eq. \eqref{Lagr1app1} by $\bm x_2^T$ to the left, and Eq. \eqref{Lagr1app2} by $\bm x_1^T$ to get
\begin{align}
\bm x_2^T A^T A\bm x_1 -\bm x_2^T A^T \bm b &=\lambda_1 \bm x_2^T \bm x_1\\
\bm x_1^T A^T A\bm x_2-\bm x_1^T A^T \bm b &=\lambda_2  \bm x_1^T \bm x_2\ .
\end{align}
Transposing the above equations, noting that $(A\bm x_2)^T(A\bm x_1)=(A\bm x_1)^T(A\bm x_2)$ and $\bm x_2^T \bm x_1=\bm x_1^T \bm x_2$ since the dot product is symmetric, and subtracting the second equation from the first, we get
\begin{equation}
 -\bm b^T A (\bm x_2-\bm x_1)=(\lambda_1-\lambda_2) \bm x_1^T \bm x_2\ .\label{sum2}
\end{equation}
Summing up Eq. \eqref{sum1} and Eq. \eqref{sum2}
\begin{equation}
||A\bm x_2||^2-||A\bm x_1||^2 -2\bm b^T A (\bm x_2-\bm x_1)=\lambda_2 ||\bm x_2||^2-\lambda_1 ||\bm x_1||^2+(\lambda_1-\lambda_2) \bm x_1^T \bm x_2\ .\label{sum3}
\end{equation}
Now, expanding Eq. \eqref{Deltadef}, we have
\begin{equation}
 \Delta(\bm x_1,\bm x_2;\lambda_1,\lambda_2)=||A\bm x_2||^2-||A\bm x_1||^2-2\bm b^T A (\bm x_2-\bm x_1)\ ,
\end{equation}
which is identical to the left hand side of Eq. \eqref{sum3}. It follows that
\begin{equation}
 \Delta(\bm x_1,\bm x_2;\lambda_1,\lambda_2)=\lambda_2 ||\bm x_2||^2-\lambda_1 ||\bm x_1||^2+(\lambda_1-\lambda_2) \bm x_1^T \bm x_2=\gamma (\lambda_2-\lambda_1)\ ,
\end{equation}
where
\begin{equation}
\gamma = N -  \bm x_1^T \bm x_2 > 0\ .
\end{equation}
Here, we have used that both vectors $\bm x_1$ and $\bm x_2$ live on the sphere of squared radius $N$, and -- if they are distinct -- their dot product must be smaller than $N$ due to Cauchy–Schwarz inequality. It follows that the cost function $H(\bm x_2)$ is larger than the cost function $H(\bm x_1)$ if the corresponding Lagrange multipliers have the same relation $(\lambda_2>\lambda_1)$, which is in essence the statement of Browne's theorem.

\section{Appendix: Calculation of the integral \eqref{intMPphi}}\label{app:integralMP}
We wish to show that
\begin{equation}
I(s_-,s_+,\lambda)=\int_{s_-}^{s_+}\mathrm{ds}~\rho_{MP}(s)\frac{s}{(\lambda-s)^2}
=\frac{1}{(\sqrt{s_+}-\sqrt{s_-})^2}\frac{(\sqrt{s_+ -\lambda}-\sqrt{s_- -\lambda})^2}{\sqrt{(s_+-\lambda)(s_- -\lambda)}}\label{toshowMP}
\end{equation}
for $\lambda<s_-$, where
\begin{equation}
\rho_{MP}(s)=\frac{2}{\pi s}\frac{1}{(\sqrt{s_+}-\sqrt{s_-})^2}\sqrt{(s_+-s)(s-s_-)}\ .
\end{equation}
Making the change of variables $t=(s-s_-)/(s_+ - s_-)$, we get
\begin{align}
I(s_-,s_+,\lambda)=\frac{2}{\pi}\frac{(s_+ - s_-)^2}{(\sqrt{s_+}-\sqrt{s_-})^2 (\lambda-s_-)^2}\int_{0}^{1}\mathrm{d}t~\frac{\sqrt{t(1-t)}}{(1+\gamma t)^2}\ ,\label{intMP}
\end{align}
where
\begin{equation}
\gamma = \frac{s_+ - s_-}{s_- -\lambda}>0\ .
\end{equation}
The integral in \eqref{intMP} admits a long but elementary antiderivative, which leads to
\begin{equation}
\int_{0}^{1}\mathrm{d}t~\frac{\sqrt{t(1-t)}}{(1+\gamma t)^2}=\frac{\pi  \left(\gamma -2 \sqrt{\gamma +1}+2\right)}{2 \gamma ^2 \sqrt{\gamma +1}}\ .\label{intMPel}
\end{equation}
Inserting \eqref{intMPel} into \eqref{intMP} and after simplifications, in order to establish \eqref{toshowMP} we need to prove that
\begin{equation}
\frac{(\sqrt{s_+ -\lambda}-\sqrt{s_- -\lambda})^2}{\sqrt{(s_+-\lambda)(s_- -\lambda)}}= \frac{ \left(\gamma -2 \sqrt{\gamma +1}+2\right)}{\sqrt{\gamma +1}}\ .\label{relation}
\end{equation}
Calling $\sqrt{s_+ -\lambda}=x$ and $\sqrt{s_- -\lambda}=y$, we have $\gamma = x^2/y^2-1$. Therefore, the relation \eqref{relation} is equivalent to
\begin{equation}
\frac{(x-y)^2}{x y}=\frac{x}{y}+\frac{y}{x}-2\ ,
\end{equation}
which is very easily verified.

\section{Appendix: Laplace method for the asymptotic evaluation of integrals}\label{appLaplace}

The Laplace method is a powerful technique used in the field of asymptotic analysis for approximating integrals. This method is particularly useful when dealing with integrals that are difficult to evaluate using standard techniques. The Laplace method is based on the principle of approximating the integral of an exponentially decaying function by the function's value at its extremal point(s).

Consider an integral of the form
\begin{equation}
    I(T) = \int_{a}^{b} \mathrm{e}^{T f(x)} g(x) \, \mathrm{d}x,
\end{equation}
where $T$ is a large parameter, and $f(x)$ and $g(x)$ are smooth functions. The objective is to find an asymptotic approximation of $I(T)$ as $T \to \infty$.

The Laplace method is based on the observation that, for large $T$, the main contribution to the integral comes from the neighbourhood of the point where $f(x)$ attains its maximum value inside the interval $(a, b)$. Assume this maximum occurs at a point $x_0 \in (a, b)$ such that $f''(x_0) < 0$. 

Expanding $f(x)$ around $x_0$ using Taylor's theorem, we get
\begin{equation}
    f(x) \approx f(x_0) + \frac{1}{2}f''(x_0)(x - x_0)^2 + \ldots\ ,
\end{equation}
where we used the fact that $f'(x_0)=0$ being a maximum. We can also similarly expand the function $g(x)$ to get
\begin{equation}
g(x)\approx g(x_0)+g'(x_0)(x-x_0)+\ldots\ .
\end{equation}

Substituting these expansions into the integral, and keeping only the first terms, we obtain
\begin{equation}
    I(T) \approx \mathrm{e}^{T f(x_0)} g(x_0)\int_{a}^{b} \mathrm{e}^{-\frac{1}{2} T |f''(x_0)|(x - x_0)^2} \left[1+\frac{g'(x_0)}{g(x_0)}(x-x_0)+\ldots\right] \, \mathrm{d}x\ .
\end{equation}
Making a change of variables $\sqrt{T}(x-x_0)=y$, we get
\begin{equation}
    I(T) \approx \mathrm{e}^{T f(x_0)} g(x_0)\int_{-\sqrt{T}(x_0-a)}^{\sqrt{T}(b-x_0)} \mathrm{e}^{-\frac{1}{2}  |f''(x_0)|y^2} \left[1+\frac{g'(x_0)}{g(x_0)}\frac{y}{\sqrt{T}}+\ldots\right] \, \mathrm{d}y\ .
\end{equation}

For large $T$, the exponential function rapidly decays away from $x_0$, allowing us to extend the limits of integration to infinity. Ignoring also the sub-leading terms in square brackets, we get to leading order for large $T$
\begin{equation}
   I(T) \approx \mathrm{e}^{T f(x_0)} g(x_0)\int_{-\infty}^{\infty} \mathrm{e}^{-\frac{1}{2}  |f''(x_0)|y^2}  \, \mathrm{d}y=\mathrm{e}^{T f(x_0)} g(x_0)\sqrt{\frac{2\pi}{T |f''(x_0)|}}\ ,
\end{equation}
where we have evaluated the Gaussian integral exactly. In Section \ref{sec:replicas2} we basically used a multivariate version of the Laplace approximation and we retained only the leading exponential behaviour.

\section{Appendix: Covariance calculation leading to Eq. \eqref{deffu}}\label{app:covariance}

In this Appendix, we evaluate explicitly the covariance $\langle V_k(\bm x^{(a)})V_\ell(\bm x^{(b)})\rangle_{A,\bm b}$ between the potential $V(\bm x)$ evaluated at two different points.

If $V_k(\bm x) = \sum_j A_{kj}x_j-b_k$ for $k=1,\ldots,M$ then 
\begin{align}
\nonumber &\small{\Big\langle\left(\sum_j A_{kj}x_j^{(a)}-b_k\right)\left(\sum_m A_{\ell m}x_m^{(b)}-b_\ell\right)\Big\rangle_{A,\bm b}}\\
\nonumber &=\sum_{j,m}x_j^{(a)}x_m^{(b)}\langle A_{kj}A_{\ell m}\rangle_A-\sum_j x_j^{(a)}\xcancel{\langle A_{kj}\rangle_A\langle b_\ell\rangle_{\bm b}}\\
 & -\sum_m x_m^{(b)}\xcancel{\langle A_{\ell m}\rangle_A\langle b_k\rangle_{\bm b}}+\langle b_k b_\ell\rangle_{\bm b}=\delta_{k\ell}f\left(\frac{\sum_m x_m^{(a)}x_m^{(b)}}{N}\right)\ ,
\end{align}
where $f(u)=\sigma^2+u$ and we used $\langle A_{ij}\rangle_A=\langle b_k\rangle_{\bm b}=0$, $\langle A_{kj}A_{\ell m}\rangle_A=(1/N)\delta_{k\ell}\delta_{jm}$ and $\langle b_k b_\ell\rangle_{\bm b}=\sigma^2\delta_{k\ell}$.

\section{Appendix: Log-Det identity (Eq. \eqref{formulalogdet})}\label{app:identities}

We give here a proof of the identity
\begin{equation}
\boxed{\frac{\partial}{\partial M_{ab}}\log\det M = [M^{-1}]_{ba}}\ .
\end{equation}
\textit{Proof:} By the chain rule
\begin{equation}
\frac{\partial}{\partial M_{ab}}\log\det M =\frac{1}{\det M}\frac{\partial}{\partial M_{ab}}\det M\ .\label{chain}
\end{equation}
Using the cofactor expansion of the determinant along the $a$-th row
\begin{equation}
\det M =\sum_{k=1}^n M_{ak}C_{ak}\ ,
\end{equation}
where the cofactor matrix $C$ is
\begin{equation}
C_{ij}=(-1)^{i+j}T_{ij}
\end{equation}
and $T_{ij}$ is a \emph{minor} of $M$, i.e. the determinant of the $(n-1)\times (n-1)$ matrix obtained removing the $i$-th row and $j$-th column of $M$. 

Hence

\begin{equation}
\frac{\partial}{\partial M_{ab}}\det M=\sum_{k=1}^n \left[\underbrace{\frac{\partial M_{ak}}{\partial M_{ab}}}_{\delta_{kb}}C_{ak}+M_{ak}\underbrace{\frac{\partial C_{ak}}{\partial M_{ab}}}_{=0}\right]\ ,\label{diffdet}
\end{equation}
where the last term vanishes as the elements in row $a$ do not affect the corresponding cofactor.

It follows from \eqref{chain} and \eqref{diffdet} that
\begin{equation}
\frac{\partial}{\partial M_{ab}}\log\det M =\frac{C_{ab}}{\det M}=\frac{[\mathrm{adj}(M)]_{ba}}{\det M}\ ,\label{finaladju}
\end{equation}
where we use the fact that the adjugate matrix $\mathrm{adj}(M)$ is the transpose of the cofactor matrix. The right hand side of \eqref{finaladju} is readily recognised as the element $ba$ of the inverse matrix of $M$.

\

\section{Appendix: Large deviations for coin tossing}\label{ref:largedev}

I will provide here a quick crash-course on large deviations for those who have not met the concept and ideas before. I strongly recommend the review \cite{touchette} by Hugo Touchette for a very thorough introduction to large deviations for physicists. Consider the following question: what is the probability of getting $M$ Heads out of $N$ (fair) coin tosses?

The exact formula for any $M,N$ is given by the following binomial distribution
\begin{equation}
P(M,N)={N\choose M}\frac{1}{2^N}\ ,
\end{equation}
where the binomial factor counts the number of possible arrangements of the $M$ Heads within the sequence of $N$ tosses.

Clearly, we can now compute the average and variance of the number $M$ of Heads as
\begin{align}
\mu &=\sum_{M=0}^N M {N\choose M}\frac{1}{2^N}=\frac{N}{2}\\
\sigma^2 &= \sum_{M=0}^N M^2 {N\choose M}\frac{1}{2^N}-\mu^2 = \frac{N}{4}\ ,
\end{align}
and appealing to Central Limit considerations, we expect that for large $N$ the probability will converge to a Gaussian centred around $\mu$ and with fluctuations given by $\sigma^2$, i.e.
\begin{equation}
P(M,N) \to \frac{1}{\sqrt{\pi (N/2)}}\exp\left[-\frac{2}{N}\left(M-\frac{N}{2}\right)^2\right]\ .\label{GaussianAppCoin}
\end{equation}
On the other hand, it is easy to compute exactly the probability of an \emph{anomalous event} characterised by \emph{all} $N$ tosses coming up Heads:
\begin{equation}
P(M=N,N) = \frac{1}{2^N}=\exp(-N \log 2)\ .\label{LargeDevCoin}
\end{equation}
Comparing the two results for large $N$ in \eqref{LargeDevCoin} and \eqref{GaussianAppCoin} clearly shows that the two formulae are inconsistent for $M\simeq N$. In other word, the very natural Gaussian fluctuation law in Eq. \eqref{GaussianAppCoin} is only valid for \emph{typical} fluctuations of the order of $\sim\sqrt{N}$ around the mean, but is inadequate to describe \emph{large} (anomalous) fluctuations where $M$ deviates from the average $N/2$ by an amount of order $N$.

Reconciling the two results requires introducing the \emph{large deviation} (or rate) function, which governs the precise way the probability distribution decays when $N$ is large, both in the vicinity and away from the most likely value. 

\begin{figure}[htb]
\begin{center}
\includegraphics[scale=0.3,clip=true]{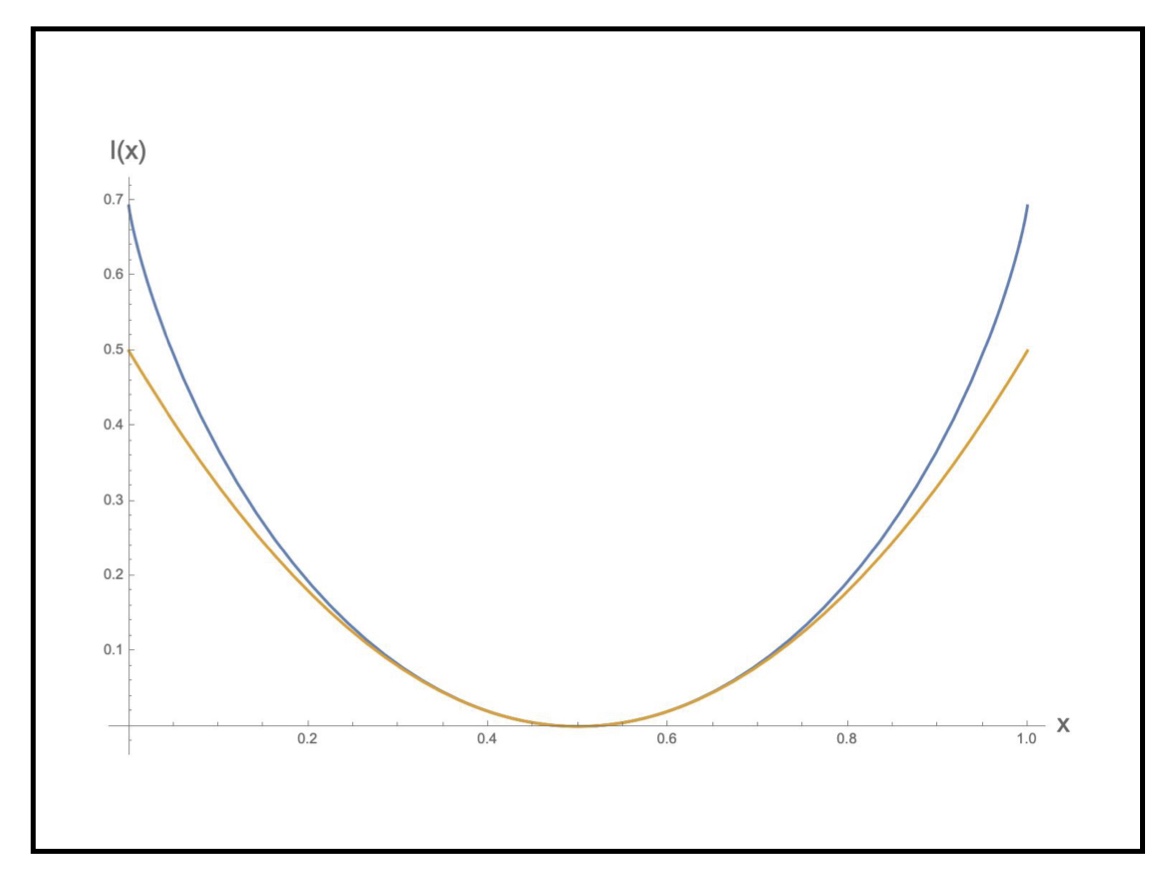}
\caption{Solid blue line: rate function $I(x)$ in Eq. \eqref{eq:rate}. Solid yellow line: quadratic behaviour $2(x-1/2)^2$ around the minimum (see Eq. \eqref{eq:quadratic}). This plot -- and the visible deviation between the two curves at the edges -- clearly shows that the Gaussian behaviour around the minimum is inadequate to characterise anomalous events characterised by a very large or small number of Heads in a series of coin tosses. } \label{rateI}
\end{center}
\end{figure}

Take $P(M=Nx,N)$ for $0\leq x\leq 1$ and expand the right hand side for large $N$ using Stirling's formula for the factorials. One obtains (neglecting pre-factors)
\begin{equation}
P(M=Nx,N) \approx \exp\left(-N I(x)\right)\ ,\label{PMlargedev}
\end{equation}
where 
\begin{equation}
I(x) = x\log x+(1-x)\log(1-x)+\log 2\ ,\label{eq:rate}
\end{equation}
where the symbol $\approx$ stands for the precise asymptotics
\begin{equation}
\lim_{N\to\infty} \frac{-1}{N}\log P(M=Nx,N)=I(x)\ .
\end{equation}
The rate function $I(x)$ is plotted in Fig. \ref{rateI}. 

For $x\to 0$ (or $x\to 1$, symmetrically) the rate function converges to the value $\log 2$, which perfectly reproduces the exact result in Eq. \eqref{LargeDevCoin}. The rate function has a minimum (a zero) at $x=1/2$, the ``most likely'' value (corresponding to $N/2$ Heads in $N$ tosses), and then increases on either side of $x=1/2$ leading $P(M=Nx,N)$ to correspondingly decay exponentially fast away from the most likely occurrence.

Interestingly enough, the rate function also provides important information about the \emph{typical} fluctuations of the random variable $M$ (= number of Heads) around its most likely value $M=N/2$. To see this, we can Taylor-expand $I(x)$ around $x=1/2$ to get
\begin{equation}
I(x) = 2 \left(x-\frac{1}{2}\right)^2+\mathcal{O}\left(\left(x-\frac{1}{2}\right)^3\right)\ .\label{eq:quadratic}
\end{equation}
Inserting this quadratic behaviour back into Eq. \eqref{PMlargedev} gives
\begin{equation}
P(M=Nx,N) \approx \exp\left(-2N  \left(\frac{M}{N}-\frac{1}{2}\right)^2\right)=\exp\left(-\frac{2}{N}  \left(M-\frac{N}{2}\right)^2\right)\ ,
\end{equation}
which precisely reproduces the Gaussian fluctuations of $M$ around its mean value $N/2$ in Eq. \eqref{GaussianAppCoin}.

The rate function $I(x)$ is therefore arguably a more fundamental and richer object than the Gaussian law \eqref{GaussianAppCoin}, as it includes the latter but provides more accurate information about larger (anomalous) fluctuations much farther away from the mean value.

%
%

\end{document}